\documentclass{ieeeaccess}
\usepackage{cite}
\usepackage{amsmath,amssymb,amsfonts}
\usepackage{algorithm}
\usepackage{algorithmic}
\usepackage{graphicx}
\usepackage{textcomp}
\def\BibTeX{{\rm B\kern-.05em{\sc i\kern-.025em b}\kern-.08em
    T\kern-.1667em\lower.7ex\hbox{E}\kern-.125emX}}
    
\usepackage{mathrsfs} 
\newlength\myindent
\newcommand\bindent{%
  \begingroup
  \setlength{\itemindent}{\myindent}
  \addtolength{\algorithmicindent}{\myindent}
}
\newcommand\eindent{\endgroup}

\begin{document}
\history{Date of publication xxxx 00, 0000, date of current version xxxx 00, 0000.}
\doi{10.1109/ACCESS.2017.DOI}

\title{Coalition Formation Games for Improved Cell-Edge User Service in Downlink NOMA and MU-MIMO Small Cell Systems}
\author{\uppercase{Panagiotis Georgakopoulos}\authorrefmark{1}, \IEEEmembership{Member, IEEE},
Tafseer Akhtar\authorrefmark{1}, \IEEEmembership{Graduate Student Member, IEEE}, Christos Mavrokefalidis\authorrefmark{2}, \IEEEmembership{Member, IEEE}, Ilias Politis\authorrefmark{1}, \IEEEmembership{Member, IEEE}, Kostas Berberidis\authorrefmark{2}, \IEEEmembership{Senior Member, IEEE} and Stavros Koulouridis\authorrefmark{1},
\IEEEmembership{Member, IEEE}}

\address[1]{Department of Electrical and Computer Engineering, University of Patras, 26504 Patras, Greece}
\address[2]{Computer Engineering and Informatics Department, University of Patras, 26504 Patras, Greece}
\tfootnote{This research has received funding in part from Project IRIDA (Cypriot RIF - INFRASTRUCTURES/1216/0017) and in part by the European Union’s Horizon 2020 Stimulating innovation by means of cross-fertilisation of knowledge program under Grant 734545 (H2020-MSCA-RISE-2016-SONNET) and Grant 872878 (H2020-MSCA-RISE-2019-eBORDER).}

\markboth
{Author \headeretal: Preparation of Papers for IEEE TRANSACTIONS and JOURNALS}
{Author \headeretal: Preparation of Papers for IEEE TRANSACTIONS and JOURNALS}

\corresp{Corresponding author: Panagiotis Georgakopoulos (e-mail: pgeorgako@ece.upatras.gr).}

\begin{abstract}
In today's wireless communication systems, the exponentially growing needs of mobile users require the combination of new and existing techniques to meet the demands for reliable and high-quality service provision. This may not always be possible, as the resources in wireless telecommunication systems are limited and a significant number of users, usually located at the cell edges, can suffer from severe interference. For this purpose, a new Joint Transmission Coordinated Multipoint (JT-CoMP) scheme, in which the transmission points' clustering is based on a coalition formation game, is deployed alongside with Non-Orthogonal Multiple Access (NOMA) in a Cloud Radio Access Network (C-RAN) consisting of small cells. To further enhance the network's performance, Multiuser Multiple-Input Multiple-Output (MU-MIMO) with Zero-Forcing (ZF) beamforming is applied. The proposed scheme's performance, in terms of user throughput, is then compared to that of a scheme where JT-CoMP with static clustering is selected, a JT-CoMP scheme with clustering based on a greedy algorithm and a scenario where JT-CoMP is not deployed. Simulation results confirm that the proposed scheme reliably improves the throughput of users with poor wireless channels while guaranteeing that the performance of the rest is not severely undermined.
\end{abstract}


\titlepgskip=-15pt

\maketitle

\section{Introduction}
\label{sec:introduction}

\IEEEPARstart{T}{he} exponential growth in mobile users' demands for increased data rates gives rise to significant challenges in terms of guaranteeing Quality of Service (QoS), increasing system capacity,  ensuring reliability \cite{b1} and robustness \cite{b2x} for Fifth generation (5G) wireless networks. To address these challenges, new and existing technologies need to coexist and cooperate in order for the emerging networks to fulfill the users' needs for high Quality of Experience (QoE) \cite{b3x}. 

An appropriate strategy to meet the ever growing demands for higher data rates and system capacity is the network densification. This is realized by densely deploying multiple low power transmission points in a given geographic area to create a network of cells with small coverage, i.e. small cells \cite{b4, b5}, while ensuring nearly uniform distribution of mobile users among them \cite{b4x}. The benefit of this approach is the reduced traffic load for every cell. Also, the mobile users and the transmission points are closer compared to cells with large coverage, i.e. macro cells, resulting in reduced path loss and increased received signal power. Cloud Radio Access Networks (C-RAN) are considered a key enabler for small cell deployment by providing efficient baseband processing over the cloud, while reducing their realization costs, enhancing their energy efficiency and reducing their power consumption \cite{b5x}. In C-RAN architecture, low-cost transmission points, called Remote Radio Heads (RRHs), are randomly and densely deployed in the network, increasing the system capacity and coverage. The RRHs are connected to the Baseband Units (BBUs), which are co-located in a virtualized BBU pool, through the fronthaul links \cite{b6}. 

One of the relatively recently proposed concepts, aiming to fulfill the needs of mobile users in 5G networks, is the Non-Orthogonal Multiple Access (NOMA) technique. NOMA has been shown to provide performance gains over Orthogonal Multiple Access (OMA) schemes, both in the downlink and the uplink \cite{b6x}, \cite{b7x}. NOMA techniques can broadly be classified into two categories, namely power domain and code domain NOMA \cite{b8x}. In power domain NOMA, simultaneous service is offered to multiple users by sharing the same time-frequency radio resources, i.e. the Resource Blocks (or RBs), among them. In a downlink scenario, the transmitter can schedule the same radio resources to multiple users, forming a NOMA cluster, by superposing their signals in the power domain and assigning different transmission power to each one of them. The superposition is done in a way where every NOMA enabled user can decode its respective incoming signal by using Successive Interference Cancellation (SIC) after reception. SIC decoding is performed according to an ascending order of the channel gains of the NOMA users, meaning that a user can eliminate the interuser interference from users with lower channel gain, while interuser interference from users with higher channel gain is treated as noise. Our study focuses only on power domain NOMA which will be referred to simply as NOMA throughout this work.

Furthermore, to increase the system's spectral efficiency Multi-user Multiple-Input Multiple-Output (or MU-MIMO) can also be implemented. MU-MIMO exploits the spatial diversity of the propagation channel, allowing parallel transmissions to multiple users \cite{b9x}. However, this induces interbeam interference to the served users, as the same RBs are employed simultaneously between users served by different beams.

MU-MIMO combined with SIC of NOMA may result in low Signal-to-Interference-plus-Noise-Ratio (SINR) especially for lower channel gain users as they receive interbeam and interuser interference. Furthermore, the dense deployment of small cells results in high intercell interference as well, as users are closer to an increased number of interfering cells, compared to macro cell systems. Therefore, these users may not achieve high data rates meaning that appropriate strategies need to be adopted in order to ensure high QoE for them. Interbeam and interuser interference can be reduced by appropriate user pairing, power allocation and beamforming \cite{b10x, b11x}, parameters that can be adjusted in NOMA and MU-MIMO.

To mitigate intercell interference, Joint Transmission Coordinated Multipoint (JT-CoMP) technique can be employed. JT-CoMP is an advanced scenario of CoMP implementation according to which each User Equipment (UE) active in the CoMP area, usually at the cell edge, is capable of receiving data simultaneously from multiple cooperating transmission points by using the same time-frequency resources, i.e. the RBs \cite{b12x}. These users are referred to as edge users in this paper.

The clustering of cooperating transmission points in JT-CoMP has been identified as a significant challenge in various studies \cite{b13, b13x,b14x}. When JT-CoMP is deployed alongside NOMA, edge users may form a NOMA cluster with non-edge users \cite{b15x}, \cite{b16x}. Regarding downlink, when the number of edge users in the network is large, incautious transmission point clustering may result in overcrowded NOMA clusters and poor data rate performance for both the edge and non-edge UEs, as the transmission power needs to be distributed amongst all of them. Therefore, the transmission point clustering needs to be intelligent to account for this cost of cooperation. Coalition formation games account for network structure and the costs of cooperation, while satisfying the individual rational demands of the network nodes, making them suitable to address this drawback \cite{b17x,b18x}.

The remainder of the paper is organized as follows. Section II presents a review of the related literature, followed by the motivation behind this study and its contributions. Section III provides an overview of the system model. In Section IV, the proposed coalition formation game is formulated and analyzed, while Section V describes the proposed coalition formation algorithm alongside its properties. In Section VI, the chosen methods for beamforming, user pairing, power allocation and multi-user scheduling are analyzed. Section VII provides the simulation setup and the simulation results are presented. Finally, Section VIII concludes the paper.

\section{Related works and contribution}
\label{Sec:Background}

\subsection{Related studies}
\label{SUBSEC:Related_Studies}
Various techniques of cooperative transmissions for interference mitigation have been studied in the literature. In \cite{b19x}, base station coordination with dirty paper coding was initially proposed with single-antenna transmitters and receivers in each cell. In \cite{b20x}, various cooperative joint transmission schemes are explored for intercell interference mitigation in a downlink multi-cell MU-MIMO network. The schemes proposed therein include a dirty paper coding approach with perfect data and power cooperation among base stations with a pooled power constraint and several sub-optimal joint transmission schemes with per-base power constraints. However, possible base station clustering schemes and large scale simulations were not included in this work, as only a three-cell scenario was considered.

More recent studies, such as \cite{b12x,b13x} and \cite{b21x}, have studied and proposed CoMP as an interference coordination scheme. Specifically, in \cite{b12x}, the JT-CoMP technique has been shown to be capable of providing the highest gains in terms of cell capacity in dense homogeneous and heterogeneous cell deployments among the various CoMP schemes. However, several challenges for its effective implementation have been identified by these studies, including the cell clustering and backhaul capacity and latency constraints.

The transmission point clustering schemes for CoMP can be generally divided in three main categories, namely static, semi-dynamic and dynamic clustering \cite{b22x}. In static clustering, the clusters are formed in a static way, usually based on the distance between the transmission points and they remain intact, regardless of network changes. Various research papers consider static clustering methods \cite{b23x,b24x,b25x}. Semi-dynamic clustering is an improved version of static clustering, with minimal overhead increase, where several layers of static clusters are designed to avoid intercluster interference. However, as in static clustering, this method of clustering is not able to adapt to user profile and network changes \cite{b26x,b27x} and usually is based on idealistic hexagonal grid topology \cite{b22x}, meaning that is not applicable in real network topologies. Therefore, fully dynamic solutions capable of being effective in scalable scenarios and capable to respond to changing network conditions to maximize CoMP gains are required.

A significant number of dynamic clustering algorithms have been proposed in the literature. In \cite{b28x}, a dynamic cluster formation algorithm is presented which merges cells into clusters based on the total improvement in spectral efficiency and users' SINR. In \cite{b29x}, the authors examine a joint dynamic base station clustering and beamformer design problem in a network where Joint Processing CoMP is deployed in the downlink, by formulating and solving a non-convex max-min rate problem. The main aim of this study is to maximize the minimum rate among all users and limit the cooperating cluster size without hurting the achievable common rate. In \cite{b30x}, a dynamic CoMP clustering algorithm is presented, aiming to maximize the group average SINR under the constraint of maximum target cell blocking probabilities for group communications in Mission-Critical Communications. The authors consider a dynamic traffic model and analyze the trade-off between high SINR and network capacity. However, in these studies, the proposed solutions are not taking into account possible individual decrease in user data rate due to resource scarcity, as the algorithms developed therein aim to improve a total performance metric and not to guarantee a performance improvement for every user individually. In \cite{b31x}, CoMP MU-MIMO downlink transmissions are examined, where the clustering is performed in multiple steps based on backhaul constraints as well as the radio properties of the network's UEs. However, the study focuses on a scenario in which each base station serves one UE, meaning that the parallel service of multiple users per cell, competing for resources, is not examined. The study showed that the CoMP gain by increasing cluster size becomes much less or even negligible if backhaul constraints are considered. In contrast with the aforementioned studies, \cite{b32x} presents a user-centric CoMP clustering algorithm for maximizing each user's spectral efficiency, meaning that each user selects its preferred set of serving cells, given a maximum cluster size. The algorithm is then enhanced to balance the load across the small cells and improve user satisfaction with a two-stage re-clustering algorithm. However, the complexity of the scheme in \cite{b32x} increases significantly with the increase of the number of users and small cells and clustering changes are proposed over longer time intervals, meaning that small-scale phenomena are not considered. The same principle is applied in \cite{b33x,b34x}, where JT-CoMP is deployed alongside NOMA in a downlink scenario. However, the simulation scenarios are fairly limited due to complexity and MIMO is not applied in the system.

A number of recent studies focus on implementing non-cooperative and cooperative game theory for the transmission point clustering in order to achieve interference mitigation in downlink transmissions, as it can provide distributed solutions with reduced signalling overhead and consider possible cooperation costs. In \cite{b35x}, a merge-only coalition formation algorithm is considered to cluster the small cell base stations so that they can perform cluster-wise beamforming to mitigate intercell interference and long-term shadow fading in a scenario with imperfect channel state information (CSI). In \cite{b36x,b37x} a coalition formation game is developed to form cooperation clusters in order to mitigate intercell interference and improve user performance via Time-Division Multiple Access (TDMA) based transmissions. However, in these studies, JT-CoMP and the parallel service of the network's users are not considered. In contrast with \cite{b32x}, a load-aware network-centric clustering coalition formation algorithm is presented in \cite{b38x}. In this study, the coalitions are formed based on merge and split operations, aiming to jointly optimize the load distribution and spectral efficiency in a JT-CoMP downlink heterogeneous network scenario. The algorithm proposed therein accounts for various overhead costs and is capable of dynamically adjusting the cluster size to adapt to different network load. As in \cite{b32x}, clustering changes are proposed over longer time intervals, meaning that small-scale phenomena are not considered. Moreover, the algorithm proposed in \cite{b38x} aims to form coalitions based on total utility improvement, ignoring possible individual payoff reductions. A study that compares static, dynamic distributed and dynamic game theoretic clustering approaches can be found in \cite{b39x}. The dynamic distributed clustering algorithm proposed therein provides a considerable improvement in terms of user throughput. However, the algorithm is not designed to account for cooperation costs. Furthermore, individual player payoff improvements or reductions are not considered in the game theoretic solution, since its objective is to increase a total utility, and JT-CoMP is not deployed.

\subsection{Motivation and contributions}
\label{SubSec:Motivation}

The aim of this work is to improve the performance, in terms of throughput, of the edge users of a 5G small cell C-RAN network where NOMA and MU-MIMO are combined and user mobility is a factor, without significantly undermining the performance of the non-edge UEs. The selected way to achieve this is via implementing JT-CoMP in which the transmission point clustering is based on a coalition formation game which takes into account the costs of cooperation and can adapt to the dynamic nature of the selected scenario. Furthermore, the obtained results are compared with those of a no-CoMP case, a case where the transmission point clustering is done in a static way and a case where a dynamic greedy clustering algorithm is applied in order to confirm the effectiveness of the proposed scheme to reliably and consistently mitigate the intercell interference and increase mobile user performance.

This paper's contributions can be summarized as follows:

\begin{enumerate}
\item To the authors' knowledge, no other study is concerned with the combined implementation of NOMA, MU-MIMO and JT-CoMP from a coalitional game theoretic perspective on a 5G small cell network consisting of low-power transmission points and moving users. Detailed equations for the received signal and SINR for each case are formulated, in addition to NOMA user cluster formation, with and without CoMP, power coefficient allocation and beamforming analysis.
\item A coalition formation game is played on behalf of the network's UEs, where the RRHs are the players, aiming to form the most beneficial JT-CoMP clusters for the UEs in the cell edges, while the performance of the rest in terms of throughput remains practically unaffected. For this purpose, appropriate utility and individual payoff functions are formulated, while a dynamic coalition formation algorithm is being developed.
\item The effectiveness of the proposed approach, in which JT-CoMP is implemented via the proposed coalition formation algorithm in a network with mobile users, is verified via simulations. The performance of the proposed scheme is then compared to that of a conventional JT-CoMP case with static clustering, a case where JT-CoMP with a greedy clustering approach has been adopted and a case where JT-CoMP is not deployed.
\end{enumerate}

Regarding notations, we will use lowercase or uppercase letters for scalars, boldface lowercase letters for vectors, boldface uppercase letters for matrices. Furthermore, the superscripts $(\cdot)^T$, $(\cdot)^H$, $(\cdot)^{-1}$, $(\cdot)^{\dagger}$ denote the transpose, the conjugate transpose, the matrix inverse and the Moore-Penrose pseudo-inverse respectively. Additionally, $|\cdot|$ and $||\cdot||$ indicate the norm and the Euclidean norm of a scalar and vector respectively.

\section{System model}

\begin{figure}[t]
\centerline{\includegraphics[width=1\columnwidth,height=\textheight,keepaspectratio]{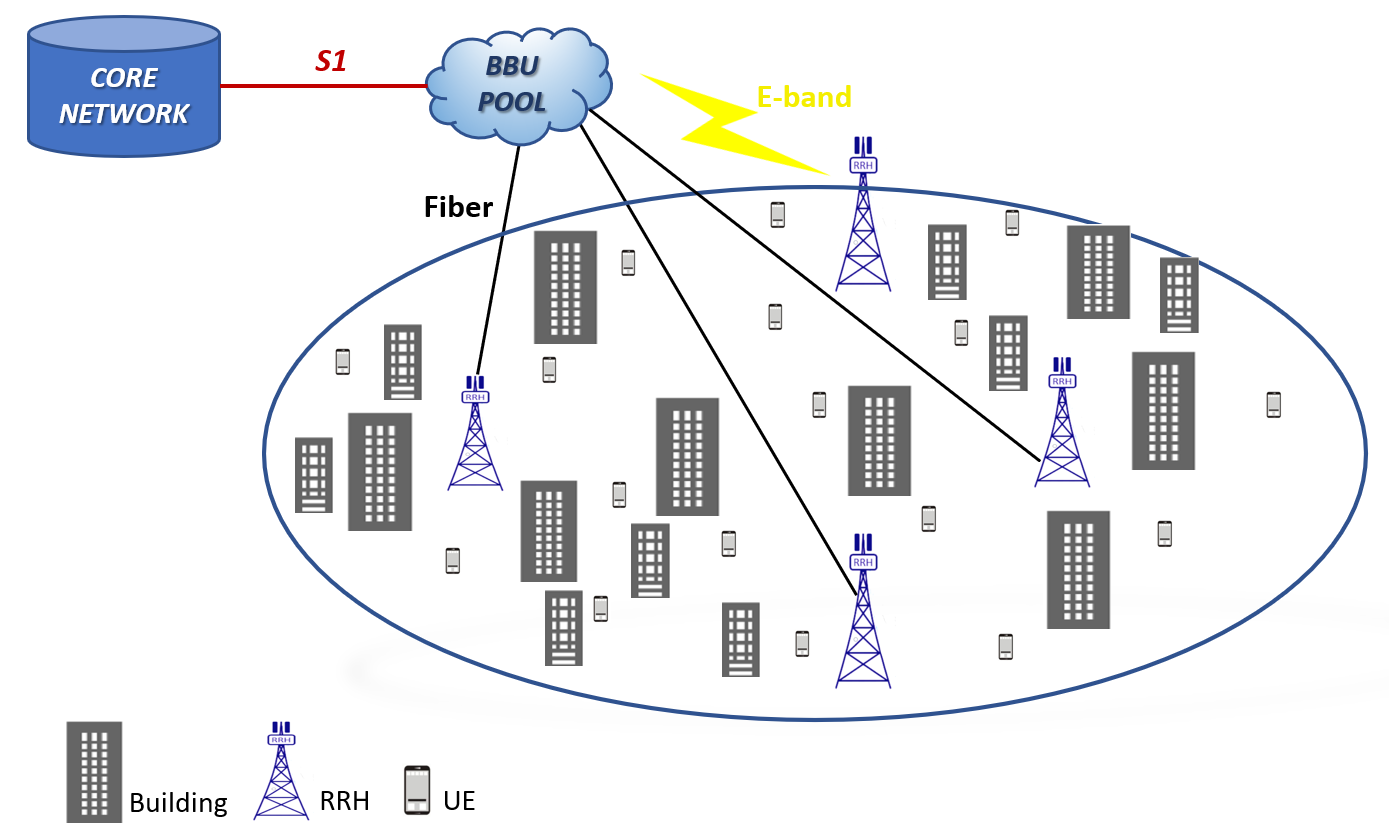}}
\caption{C-RAN small cell network architecture.}
\label{fig1}
\end{figure}

A downlink scenario in a small cell C-RAN based network for a hyperdense urban environment is considered, as shown in Fig.~\ref{fig1}. Unlike traditional architectures, the base station is separated into a radio unit, i.e. the RRH, and a signal processing unit, i.e. the BBU. As a result, the network consists of three main parts, the backhaul, the fronthaul and the access network. The backhaul connects the virtualized BBU pool, where BBUs from multiple sites are co-located, with the mobile core network over the S1 interface. The advantage of this approach is that the baseband processing is centralized and shared among multiple sites, resulting in more efficient utilization of the available resources, reduced delay and network costs of operation \cite{b40x}. The fronthaul part of the network spans from the RRHs to the BBU pool. The RRHs are connected to the high performance processors of the virtualized BBU pool over wireless microwave links in the E-band or optical fiber using the Common Public Radio Interface (CPRI) protocol \cite{b40x,b41x}. The interface which connects the RRHs with the BBU pool is defined as the Ir interface. Finally, in the access network, multiple densely deployed RRHs serve the network devices, e.g. the UEs.

In this study, the access network consists of low-power RRHs equipped with \(N\) antennas each as the transmission points, serving multiple moving single antenna UEs, randomly distributed in the RRH coverage area. It is assumed that the frequency reuse factor in the network is one. Cyclic Prefix Orthogonal Frequency-Division Multiplexing (CP-OFDM) is utilized for the formation of the transmitted signals, as in 5G New Radio (NR) \cite{b42x}. Depending on its available bandwidth, each RRH has a fixed number of available RBs to allocate to its users per Transmission Time Interval (TTI), where each RB consists of \(F=12\) contiguous and orthogonal subcarriers\cite{b42x} and each TTI has a duration of \(O\) OFDM symbols.
Let us assume a total of \(L\) RRHs, forming \(L\) total cells, where $\mathcal{L}$ = \{RRH$_{1}$, RRH$_{2}$,\dots, RRH$_{l}$,\dots, RRH$_{L}$\} represents the set of RRHs in the network. NOMA is the selected multiple access scheme, meaning that, in each cell, each RB can be allocated to a group of UEs which is referred to as a NOMA cluster.

Considering MU-MIMO as the underlying technology, a total of \(N\) beams can be generated by each RRH and each beam \(n\) can serve the users of one NOMA cluster. Let $\mathcal{K}_{l,n}$ = \{UE$_{l,n,1}$, UE$_{l,n,2}$,\dots, UE$_{l,n,k}$,\dots, UE$_{l,n,K}$\} denote the set of UEs in a NOMA cluster served by RRH$_{l}$ and beam \(n\) for an allocated RB, where \(K \geq 2\) . Therefore, each beam serves \(K\) users in total via NOMA, while every RRH transmits \(N\) total beams, meaning that transmission to a maximum number of \(K \cdot N\) UEs in parallel is possible in the same RB. The superimposed signal to be transmitted at the subcarrier \(f\) and OFDM symbol \(o\), without including these indices from here on for the sake of simplicity, in the \(n\)-th beam of RRH$_{l}$ can be expressed as:

\begin{equation}
x_{l,n}=\sum_{k=1}^{K}\sqrt{a_{l,n,k}P_{l,n}}s_{l,n,k} ,
\end{equation}
where \(P_{l,n}\) is the total transmit power allocated to beam \(n\) from RRH$_{l}$ for the subcarrier \(f\) and OFDM symbol \(o\) and \(a_{l,n,k}\) denotes the power coefficient which is the fraction of \(P_{l,n}\) that is allocated to user UE$_{l,n,k}$ according to NOMA. For every NOMA cluster it holds that $\sum_{k=1}^{K}a_{l,n,k} = 1$. Also, it is assumed that the total transmit power of each RRH is constant over time and is divided equally among all beams and available subcarriers. Finally, \(s_{l,n,k}\) is the desired signal of UE$_{l,n,k}$ and is assumed that $\mathbb{E}(|s_{l,n,k}|^2) = 1$.
The received signal of UE$_{l,n,k}$ at the subcarrier \(f\) and OFDM symbol \(o\), served by RRH$_{l}$ and beam \(n\) is described by the following equation:

\begin{equation}\resizebox{0.98\hsize}{!}{$
\begin{aligned}
y_{l,n,k} & = \displaystyle\sum_{l=1}^{L}\displaystyle\sum_{n=1}^{N}\displaystyle\sum_{k=1}^{K} \boldsymbol{h}_{l,n,k}\boldsymbol{w}_{l,n}\sqrt{a_{l,n,k}P_{l,n}}s_{l,n,k} + z_{k}\\
&=\boldsymbol{h}_{l,n,k}\boldsymbol{w}_{l,n}\sqrt{a_{l,n,k}P_{l,n}}s_{l,n,k} \\
 & + \displaystyle\sum_{k'=1,\neq k}^{K}\boldsymbol{h}_{l,n,k}\boldsymbol{w}_{l,n}\sqrt{a_{l,n,k'}P_{l,n}}s_{l,n,k'} \\
 & + \displaystyle\sum_{n'=1,\neq n}^{N}\displaystyle\sum_{k'=1}^{K} \boldsymbol{h}_{l,n,k}\boldsymbol{w}_{l,n'}\sqrt{a_{l,n',k'}P_{l,n'}}s_{l,n',k'} \\
 & + \displaystyle\sum_{l'=1,\neq l}^{L}\displaystyle\sum_{n'=1}^{N}\displaystyle\sum_{k'=1}^{K} \boldsymbol{h}_{l',n',k}\boldsymbol{w}_{l',n'}\sqrt{a_{l',n',k'}P_{l',n'}}s_{l',n',k'} \\
 & + z_{k} ,
\end{aligned}$}
\end{equation}
where \(\boldsymbol{w}_{l,n} \in\) $\mathbb{C}^{N\times 1}$ is the beamforming vector for beam \(n\) of RRH$_{l}$ and ${z}_{k}\ \mathrm{\sim }\mathrm{\ }\mathcal{N}_{\mathbb{C}}\left(0,{\sigma }^2\right)$ is the Additive White Gaussian Noise (AWGN) at UE$_{l,n,k}$. The first term of (2) is the useful signal at the UE, the second one represents the intrabeam interference, i.e interference from UEs served in the same beam, the third is the interbeam interference, i.e interference from other beams of the serving RRH, and finally the fourth term is the intercell interference, i.e interference from the beams of other RRHs in the network. Finally, \(\boldsymbol{h}_{l,n,k} \in\) $\mathbb{C}^{1\times N}$ is the channel gain vector between UE$_{l,n,k}$ and the antennas of RRH$_{l}$ and is given by the following equations:

\begin{equation}
\boldsymbol{h}_{l,n,k}=\sqrt{10^{-v_{l,n,k}/10}}\boldsymbol{r} ,
\end{equation}

\begin{equation}
v_{l,n,k}=\Lambda_{l,n,k}+S-G_{T}-G_{R} ,
\end{equation}
where \(\Lambda_{l,n,k}\) is the path loss between RRH$_l$ and UE$_{l,n,k}$ in dB, \(S\) is the shadowing coefficient in dB which follows a zero mean normal distribution with a propagation topology specific shadowing standard deviation. Also, \(G_{T}\) and \(G_{R}\) symbolize the total antenna gain of RRH$_{l}$ and the antenna gain of the UE$_{l,n,k}$, respectively, in dBi. We assume that all the RRHs in the network have the same total antenna gain, while the same assumption is made for the antenna gains of the network's UEs. Finally, \(\boldsymbol{r}\) is the Rayleigh small-scale fading coefficient which is represented as an \(1 \times N\) complex number vector with independent and normally distributed components with zero mean value and unit variance.

In NOMA, users in the same NOMA cluster share the same RB, resulting in interuser interference, which is referred to as intrabeam interference in this study, as MU-MIMO is also deployed. To deal with this type of interference, SIC is employed. In SIC, the received superimposed signal for the NOMA cluster is decoded in the ascending order of the cluster's users' channel gains. Assuming that the channel gains for UEs in the same NOMA cluster are sorted as \(\|\boldsymbol{h}_{l,n,1}\|\leq \|\boldsymbol{h}_{l,n,2}\|\leq \dotso \leq \|\boldsymbol{h}_{l,n,K}\|\), UE$_{l,n,k}$ will decode the signal of UE$_{l,n,i}$ and remove the interuser interference caused by it if \(k>i\). If \(k<i\), then the signal of UE$_{l,n,i}$ is treated as interference \cite{b43x}.
Thus, after SIC is applied, the per-subcarrier and per-symbol SINR of UE$_{l,n,k}$ is expressed as follows:

\begin{equation}
SINR_{l,n,k}=\frac{|\boldsymbol{h}_{l,n,k}\boldsymbol{w}_{l,n}|^{2}a_{l,n,k}P_{l,n}}{I_{intrabeam}+I_{interbeam}+I_{intercell}+\sigma^{2}} ,
\end{equation}
where:
\begin{equation}
I_{intrabeam} = \displaystyle\sum_{k'=k+1}^{K}|\boldsymbol{h}_{l,n,k}\boldsymbol{w}_{l,n}|^{2}a_{l,n,k'}P_{l,n} ,
\end{equation}

\begin{equation}
I_{interbeam} = \displaystyle\sum_{n'=1,\neq n}^{N}|\boldsymbol{h}_{l,n,k}\boldsymbol{w}_{l,n'}|^{2}P_{l,n'} ,
\end{equation}
and
\begin{equation}
I_{intercell} = \displaystyle\sum_{l'=1, \neq l }^{L}\displaystyle\sum_{n'=1}^{N}\displaystyle|\boldsymbol{h}_{l',n',k}\boldsymbol{w}_{l',n'}|^{2}P_{l',n'} ,
\end{equation}
where $\sigma^{2}$ is the noise power and is assumed to be equal at all receivers. Additional assumptions have been made regarding the sufficient capacity of all links between the BBU pool and the RRHs, as well as the availability of perfect CSI at the BBU pool which is reported by the UEs with zero delay. Plenty of works study NOMA with imperfect CSI such as \cite{b44x} and \cite{b45x}. However, perfect CSI is assumed in this work to study the performance upper bound of the examined system as it allows perfect SIC application and optimal resource allocation. This assumption is commonly adopted in relevant literature \cite{b46x}, \cite{b47x}. All the subcarriers of an RB experience flat-fading and all receivers treat interference as noise. Furthermore, the channel gain vectors are assumed constant for the duration of a TTI and for all the subcarriers of an RB, while all the channels of different links are assumed uncorrelated between different TTIs and independent of each other. Also, the duration of a TTI is assumed equal to the duration of a 5G NR slot and depends on the selected numerology. This means that equations (1)-(8) are applicable for the RB and slot level.

To obtain the throughput of a user UE$_{l,n,k}$, the Transport Block (TB) size needs to be determined. According to \cite{b48x}, the TB size is dependant on the total number of allocated RBs for the user, the number of subcarriers in an RB, the number of OFDM symbols of the Physical Downlink Shared Channel (PDSCH) allocation within the slot, the number of Resource Elements (REs) for Demodulation Reference Signals (DM-RS) in the scheduled duration, the number of overhead REs, the number of MIMO layers and the selected Modulation and Coding Scheme (MCS). In NOMA, the power allocation and the users in NOMA clusters are determined per-RB basis, meaning that the wideband selection of MCS is not capable of fully exploiting the benefits of NOMA \cite{b43x}.
Therefore, in this study, an MCS is selected separately for each RB. To determine the appropriate MCS for a user, the SINR for each RB that the user is scheduled is mapped to a Channel Quality Indicator (CQI) value according to \cite{b49x} for a target Block Error Rate (BLER) of 0.1. Then, an individual TB size is calculated as in \cite{b48x} for each RB scheduled for the user. Finally, the total TB size is determined by adding all the individual values, and used to extract, according to the selected numerology, the instantaneous throughput \(\delta_{l,n,k}\) of the user UE$_{l,n,k}$ in a TTI.

To mitigate intercell interference, the transmission points, i.e. the RRHs, are capable of forming cooperating clusters, also referred to as coalitions, to coordinate their transmissions via JT-CoMP. Let $\mathcal{P}$ = \{$\Pi_{1}$, $\Pi_{2}$,\dots, $\Pi_{i}$,\dots, $\Pi_{I}$\} denote the set of coalitions in the network. When a coalition $\Pi_{i}$ is formed, JT-CoMP is activated and used to transmit in the same RBs from the different clustered RRHs to users with poor wireless links. We refer to these users as edge users and the rest as non-edge users. In this study, this discrimination is based on the individual effective SINR value of each user which is obtained by averaging the SINR values for all the RBs that the user is scheduled in a TTI.

To implement JT-CoMP in NOMA, the same conditions stated at \cite{b15x} were adopted. Specifically, the signals of edge UEs are going to be decoded prior to those of the non-edge UEs, while the decoding order for an edge UE will be the same in all NOMA clusters that serve it. In our case, where MU-MIMO is also implemented, the latter condition implies that alongside their original serving beam, the edge UEs will also be assigned to the preexisting beams of the other RRHs that participate in the coalition. Also, with JT-CoMP, it is possible for a beam to serve multiple edge UEs, served originally by different RRHs before JT-CoMP was activated. Therefore, the maximum number of UEs that a beam serves at each RB after JT-CoMP activation may be greater than \(K\) and is dependant on the beam assignment for the edge UEs. 

\begin{figure}[t!]
\centerline{\includegraphics[width=1\columnwidth,height=\textheight,keepaspectratio]{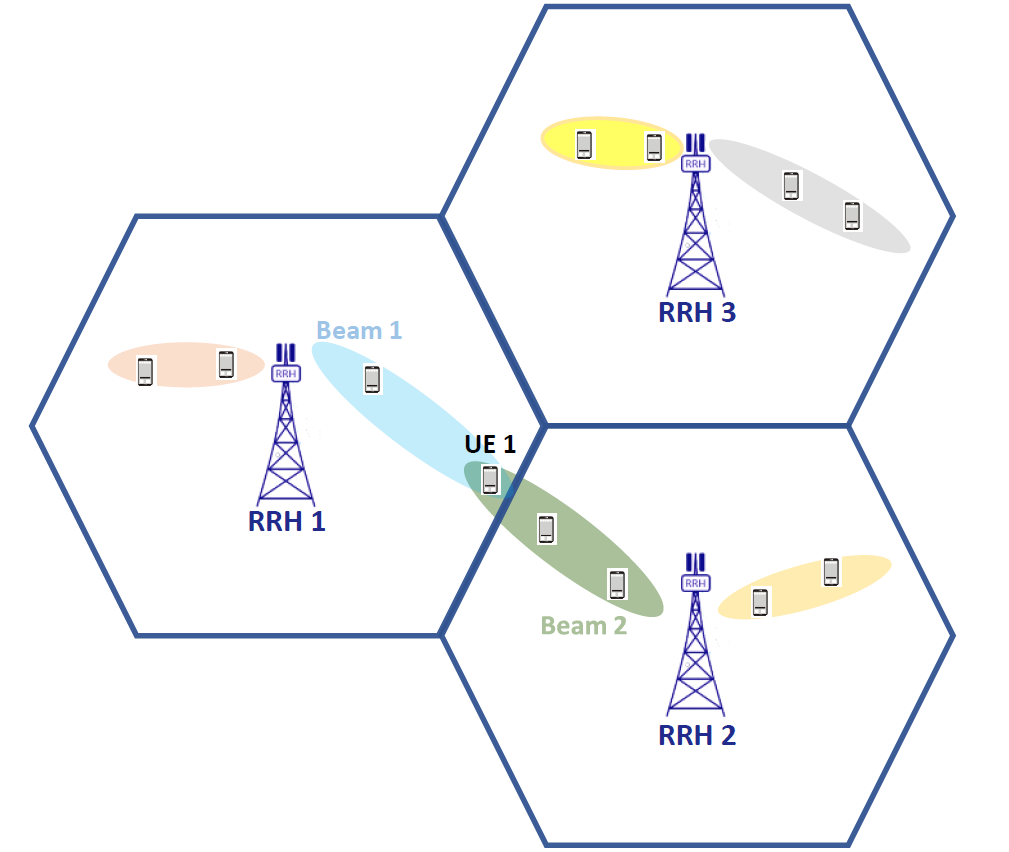}}
\caption{JT-CoMP in a three-cell NOMA MU-MIMO system.}
\label{fig2}
\end{figure}

An example of JT-CoMP implementation in our scenario for an edge user is depicted in Fig.~\ref{fig2}. In this example, each RRH can generate two beams to serve its respective NOMA clusters, which consist of two UEs each before JT-CoMP is applied. To serve potential edge users, RRH 1 and RRH 2 form a coalition, while RRH 3 is assumed to not be a part of a coalition. When edge user UE 1, served by RRH 1 before JT-CoMP is applied, is scheduled at an RB, RRH 2 needs to assign a beam to the user. In this example, based on the user pairing methodology, RRH 2 assigns Beam 2 to the edge user. By doing so, Beam 2 now serves three total UEs in its assigned NOMA cluster, whilst Beam 1 continues to serve two. Finally, UE 1 is going to be decoded first in both serving NOMA clusters as per \cite{b15x}.

The per-subcarrier and per-OFDM symbol SINR of an edge user UE$_{l,n,k_e}$\(\in\) $\mathcal{K}_{l,n}$, after forming ${\Pi}_i$, activating JT-CoMP and applying SIC, is given by the following equation:

\begin{equation}
SINR_{l,n,k_{e}}=\frac{\displaystyle\sum_{j\in \Pi_{i}}|\boldsymbol{h}_{j,n,k_{e}}\boldsymbol{w}_{j,n}|^{2}a_{j,n,k_{e}}P_{j,n}}{I_{intrabeam}+I_{interbeam}+I_{intercell}+\sigma^{2}} ,
\end{equation}
where 
\begin{equation}
I_{intrabeam} = \displaystyle\sum_{k'=k_{e}+1}^{K'}|\boldsymbol{h}_{l,n,k_{e}}\boldsymbol{w}_{l,n}|^{2}a_{l,n,k'}P_{l,n} ,
\end{equation}
\begin{equation}
I_{interbeam} = \displaystyle\sum_{n'=1,\neq n}^{N}|\boldsymbol{h}_{l,n,k_{e}}\boldsymbol{w}_{l,n'}|^{2}P_{l,n'} ,
\end{equation}
and
\begin{equation}
\begin{split}
I_{intercell} & = \displaystyle\sum_{l' \notin \Pi_{i} }\displaystyle\sum_{n'=1}^{N}\displaystyle|\boldsymbol{h}_{l',n',k_{e}}\boldsymbol{w}_{l',n'}|^{2}P_{l',n'} \\
& + \displaystyle\sum_{j \in \Pi_{i}\setminus l}\displaystyle\sum_{k'=k_e+1}^{K'} |\boldsymbol{h}_{j,n,k_{e}}\boldsymbol{w}_{j,n}|^{2}a_{j,n,k'}P_{j,n} \\
& + \displaystyle\sum_{j\in \Pi_{i}\setminus l}\displaystyle\sum_{n'=1,\neq n}^{N}|\boldsymbol{h}_{j,n',k_{e}}\boldsymbol{w}_{j,n'}|^{2}P_{j,n'} ,
\end{split}
\end{equation}
where \(K'\) is the number of users in the \(n\)-th NOMA cluster of an RRH after JT-CoMP is applied in the system. It is assumed that the per-subcarrier and per-symbol allocated power \(P_{l,n}\) to the \(n\)-th beam of RRH$_{l}$ is equal for both JT-CoMP and non JT-CoMP cases. Equations (9) and (12) indicate that by applying JT-CoMP, a part of the allocated transmit power for NOMA clusters of interfering RRHs in $\Pi_{i}$ is used for transmitting useful signal to UE$_{l,n,k_e}$, resulting in decreased intercell interference and additional received signal power for the edge user. For a non-edge user UE$_{l,n,k_{ne}}$\(\in\) $\mathcal{K}_{l,n}$, its per-subcarrier and per-symbol SINR is described by equations (5)-(8), with the difference that the number of users in its assigned NOMA cluster is equal to \(K'\), due to JT-CoMP implementation.

The instantaneous throughputs \(\delta_{l,n,k_{e}}\) and \(\delta_{l,n,k_{ne}}\) of an edge user UE$_{l,n,k_e}$ and non-edge user UE$_{l,n,k_{ne}}$ respectively, are calculated as in the case that JT-CoMP is not implemented in the system. 

\section{A new game theoretic perspective}

In this section, the possible drawbacks of JT-CoMP implementation in NOMA and MU-MIMO systems will be analyzed, followed by the reasons why coalition formation games are an attractive approach for their solution. Finally, the formulation of the proposed coalition formation game will be provided, along with various definitions about coalitional game theory.

In network densification scenarios with small cells, the distance between interfering transmission points is significantly closer compared to macro cell scenarios for the same topology, while their number is also increased. This means that in MU-MIMO and NOMA small cell systems, such as the one presented in this study, besides intrabeam and interbeam interference, the UEs can often suffer from severer intercell interference, especially if they are located in cell edges. The objective of this study is to increase the performance of these UEs, in terms of throughput, by implementing JT-CoMP. However, the implementation of JT-CoMP in NOMA and MU-MIMO systems carries some drawbacks.

When JT-CoMP is activated and coalitions of transmission points are formed, each edge UE served by an RRH participating in a coalition, is also assigned to NOMA clusters of the coalition's cooperating RRHs, besides its original serving NOMA cluster before JT-CoMP was applied \cite{b15x}. Depending on the beam assignment for the edge UEs, the number of users in some NOMA clusters will increase compared to the case that JT-CoMP is not activated and the allocated transmit power for the beams of these NOMA clusters will be distributed to more users. This means that the allocated power for the non-edge users in these NOMA clusters will decrease, while their intrabeam interference might increase, depending on their decoding order, resulting in lower SINR and throughput for these users.

Also, according to equations (5)-(8) and (9)-(12), if \(K'\) is greater than \(K\) for the original serving NOMA cluster of UE$_{l,n,k_e}$, its allocated power from this NOMA cluster will decrease, while its intrabeam interference might increase with JT-CoMP, depending on its decoding order. Also, if the other NOMA clusters to which UE$_{l,n,k_e}$ is assigned are overcrowded with users, its total allocated power might end up lower when JT-CoMP is activated and could result in decreased SINR and throughput. Therefore, JT-CoMP might undermine the performance of the edge UEs as well, despite reducing their received intercell interference.

Despite these drawbacks, the RRHs have a strong incentive to organize themselves into coalitions to implement JT-CoMP. The more RRHs participate in a coalition, the more the received power for an edge user increases, while its intercell interference decreases, resulting in increased throughput, as long as its assigned NOMA clusters are not overcrowded. Also, alongside distance, shadowing and small-scale fading determine a UE's channel gain, meaning that forming coalitions with the RRHs closer to each other will not necessarily increase their edge UEs' throughput. Furthermore, in our scenario, users are moving, meaning that mobility and varying channels cause constant changes in user profiles (a user might change its status to edge while he was non-edge and vice versa) and activate handovers, changing the load of the RRHs. This may also lead to the formation of more overcrowded NOMA clusters. Therefore, intelligence needs to be behind the JT-CoMP transmission point clustering to avoid overcrowded NOMA clusters, which is why a \((\mathcal{L},u)\) coalition formation game was formulated to cluster the network's RRHs into coalitions, where \(\mathcal{L}\) presents the set of players, while \(u\) is the utility function or value of a coalition. Coalition formation games account for costs of cooperation, like the possible decrease of the UEs' throughput when overcrowded NOMA clusters are formed, while enabling the game's players to make individual rational strategic decisions \cite{b17x,b18x}, rendering them an attractive approach to tackle cooperation drawbacks.

\textbf{Definition 1}: A coalitional structure $\mathcal{P}$ is defined as a partition of \(\mathcal{L}\) when for a collection of coalitions $\mathcal{P}$ = \{$\Pi_{1}$, $\Pi_{2}$,\dots, $\Pi_{i}$,\dots, $\Pi_{I}$\}, it holds that \(\forall i\neq j\), \(\Pi_{i}\cap \Pi_{j}=\emptyset\) and \(\bigcup_{i=1}^{I}\Pi_{i}=\mathcal{L}\) \cite{b18x}.

In this work, the focus is to increase the throughput of the system's edge users without compromising the throughput of the non-edge users significantly.
Although JT-CoMP benefits the individual throughput of edge UEs in the network, its implementation demands the formation of coalitions consisting of RRHs. This means that the players in the proposed coalition formation game are the RRHs and a function that captures the throughput changes for each user of an RRH when it is participating in a coalition \(\Pi_{i}\) has to be formulated. A suitable function for this purpose, referred to as payoff function of RRH$_{l}$\(\in\)\(\Pi_{i}\), is given by:

\begin{equation}
\begin{split}
\phi_{l}(\Pi_{i})&= \displaystyle\sum_{k_{e}=1}^{K_{l,e}}sgn(\delta_{l,k_{e}}^b-\delta_{l,k_{e}}^a)\\
&+\sum_{k_{ne}=1}^{K_{l,ne}}sgn(\delta_{l,k_{ne}}^b-(1-d_{f})\delta_{l,k_{ne}}^a)\\
&-(K_{l,e}-1-q_{l,e}+\xi_{l,e})\\
&-(K_{l,ne}-1-q_{l,ne}+\xi_{l,ne})
+\phi_{l}^a ,
\end{split}
\end{equation}
where \(d_{f}\) represents the acceptable decrease in non-edge UE throughput, \(K_{l,e}\) and \(K_{l,ne}\) are the total number of edge and non-edge UEs served by RRH$_{l}$ respectively and $\delta_{l,k_{e}}$ and $\delta_{l,k_{ne}}$ are their throughput.
The notation \(b\) corresponds to user throughput values after coalition \(\Pi_{i}\) is formed, while the notation \(a\) refers to partitions that satisfied the set conditions. However, \(\delta_{l,k_{ne}}^a\) represents always the throughput values of a non-edge UE$_{l,k_{ne}}$ without JT-CoMP's impact. Also, \(q_{l,e}\) and \(\xi_{l,e}\) are the number of edge users of RRH$_{l}$ whose throughput was unchanged or decreased by testing \(\Pi_{i}\). Similarly, \(q_{l,ne}\) and \(\xi_{l,ne}\) are the number of non-edge users of RRH$_{l}$ whose throughput values are equal to or decreased compared to their individual throughput thresholds by testing \(\Pi_{i}\).
To account for the rare cases that before and after a coalition is examined edge users had the exact same throughput and non-edge users had their throughput become equal to their threshold, the sign function was used in (13). The sign function for two throughput values \(\delta^b\) and \(\delta^a\) can be expressed as:

\begin{equation}
sgn(\delta^b-\delta^a)=\left\{
\begin{array}{ll}
      -1 &, \delta^b<\delta^a \\
      0 &, \delta^b=\delta^a \\
      1 &, \delta^b>\delta^a \\
\end{array} .
\right.
\end{equation}

When at least one condition is not satisfied, i.e. at least one edge UE's throughput decreased or one non-edge UE's throughput decreased below its corresponding threshold, while examining a coalition, the summation of the first four terms of (13) yields a negative value, resulting in a decrease of the RRH's individual payoff which indicates that the examined coalition is not beneficial for at least one of its UEs. If both conditions are satisfied, the payoff of the RRH increases, indicating that the coalition is beneficial for the throughput of its edge UEs while the throughput of its non-edge UEs is not severely undermined. Therefore, the game can be played on behalf of the UEs, while the RRHs are considered the players.

Finally, we define the utility function of coalition \(\Pi_{i}\) as:
\begin{equation}
u(\Pi_{i})= \displaystyle\sum_{l\in \Pi_{i}}\phi_{l}(\Pi_{i}) .
\end{equation}

More specifically, the formulated game is a coalition formation game with non-transferable utility (NTU), defined by a pair \((\mathcal{L},u)\), where \(\mathcal{L}\) presents the finite set of players and \(u\) as a characteristic function which associates with every coalition \(\Pi_{i}\subseteq \mathcal{L}\) a set of payoff vectors.

\textbf{Definition 2}: A coalition formation game is said to have a non-transferable utility if the value of coalition \(\Pi_{i}\), \(u(\Pi_{i})\), cannot be arbitrarily apportioned amongst its members. Instead, each member of \(\Pi_{i}\) has its own utility within a coalition \cite{b50x}.

\textbf{Property 1}: The proposed coalition formation game is an NTU game.

\emph{Proof}:
In an NTU game, the payoff that each player in a coalition \(\Pi_{i}\) receives is dependent on the joint actions that the members of \(\Pi_{i}\) select \cite{b18x}. This is the case for the examined scenario, as depending on the user scheduling and the beam assignment for the edge UEs, each player's payoff in \(\Pi_{i}\) can be obtained by (13) and not arbitrarily allotted by \(u(\Pi_{i})\) \cite{b18x}.

\textbf{Definition 3}: A coalition formation game \((\mathcal{L},u)\) is in characteristic form if the value of a coalition \(\Pi_{i}\) depends solely on the members of that coalition, with no dependence
on how the players in $\mathcal{L}$\(\setminus \Pi_{i}\) are structured \cite{b18x}.

\textbf{Property 2}: The proposed coalition formation game is in characteristic form.

\emph{Proof}: As indicated by equations (5)-(8) and (9)-(12), the per-subcarrier and per-OFDM symbol SINR for edge and non-edge UEs inside a coalition \(\Pi_{i}\), whether JT-CoMP is activated or not, depends on their decoding order and the power allocated to them and the other UEs in their serving NOMA cluster(s). Also, it is indicated that the interference from RRHs outside \(\Pi_{i}\) is expressed by the same terms for both non JT-CoMP and JT-CoMP cases, regardless of their structure. According to (13), the payoff of each player in \(\Pi_{i}\) is a function of its edge and non-edge UEs' throughput which, in turn, is a function of their per-subcarrier and per-OFDM symbol SINR values. This means that the value of a coalition \(\Pi_{i}\), given by (15), depends solely on the members of that coalition, with no dependence on how the players in $\mathcal{L}$\(\setminus \Pi_{i}\) are structured, thus, the game is in characteristic form.
 
\textbf{Definition 4}: A coalition formation game \((\mathcal{L},u)\) is said to be superadditive if for any two disjoint
coalitions \(\Pi_{c},\Pi_{d}\subset \mathcal{L}\), \(u(\Pi_{c} \cup \Pi_{d})\geq u(\Pi_{c}) + u(\Pi_{d})\) \cite{b18x}.

\textbf{Property 3}: The proposed coalition formation game is, in general, non-superadditive.

\emph{Proof}:
For two disjoint coalitions \(\Pi_{c}\subset \mathcal{L}\) and \(\Pi_{d}\subset \mathcal{L}\), \(u(\Pi_{c} \cup \Pi_{d})\) may not always be greater than \(u(\Pi_{c}) + u(\Pi_{d})\) due to possible non-edge UEs' throughput reduction below their threshold or possible edge UEs' throughput reduction which can result in decreased payoff values for the involved RRHs and decreased utility when \(\Pi_{c} \cup \Pi_{d}\). Therefore, the formulated game is non-superadditive and the coalition of all the players in the game, i.e. the grand coalition, is not the optimal structure.

\section{Proposed coalition formation algorithm}

In this section, the proposed coalition formation \textbf{Algorithm 1} alongside its various properties will be presented, after some useful definitions are provided. The proposed coalition formation game algorithm is based on the algorithm of \cite{b51x,b52x} and \cite{b53x}. In this study however, similarly to \cite{b53x}, an extra condition on the maximum coalition size was adopted to guarantee limited feedback overhead, alongside two external rules about its reactivation due to user mobility and the time-varying nature of the wireless channels. These external rules will be explained later in this Section. The proposed algorithm is based on forming coalitions of transmission points by comparing different collections of coalitions.

\textbf{Definition 5}: A comparison operator $\triangleright$ is defined for comparing two collections of coalitions $\mathcal{R}$ = \{$R_{1}$, $R_{2}$,\dots, $R_{r}$\} and $\mathcal{T}$= \{$T_{1}$, $T_{2}$,\dots, $T_{t}$\} that are partitions of the same subset \(\mathcal{B}\subseteq \mathcal{L}\) (same players in $\mathcal{R}$ and $\mathcal{T}$). Therefore, $\mathcal{R}$ $\triangleright$ $\mathcal{T}$ implies that the way $\mathcal{R}$ partitions $\mathcal{B}$ is preferred to the way $\mathcal{T}$ partitions $\mathcal{B}$ \cite{b54x}.

The criterion based on which we compare two collections is the Pareto order \cite{b54x}.

\textbf{Definition 6}: Based on Pareto order $\mathcal{R}$ $\triangleright$ $\mathcal{T}$ if \(\phi_{j}(\mathcal{R}) \geq \phi_{j}(\mathcal{T}), \forall j\in \mathcal{R},\mathcal{T}\) with at least one strict inequality (\(>\)) for a player \(j\) \cite{b54x}.

The proposed algorithm operates on two main rules, referred to as merge and split rules, which aim to merge or split coalitions based on the Pareto order.

\textbf{Definition 7}: A set of coalitions \{$\Pi_{1}$, $\Pi_{2}$,\dots, $\Pi_{J}$\} will merge if \(\{\bigcup_{j=1}^{J}\Pi_{j}\}\triangleright \{\Pi_{1}, \Pi_{2},\dots, \Pi_{J}\}\), therefore, \( \{\Pi_{1}, \Pi_{2},\dots, \Pi_{J}\}\rightarrow\{\bigcup_{j=1}^{J}\Pi_{j}\}\), \( \{\Pi_{1}, \Pi_{2},\dots, \Pi_{J}\}\subseteq \mathcal{P}\).

\textbf{Definition 8}: A coalition \(\bigcup_{j=1}^{J}\Pi_{j}\) will split if \(\{\Pi_{1}, \Pi_{2},\dots, \Pi_{J}\}\triangleright\{\bigcup_{j=1}^{J}\Pi_{j}\}\), therefore, \(\{\bigcup_{j=1}^{J}\Pi_{j}\}\rightarrow\{\Pi_{1}, \Pi_{2},\dots, \Pi_{J}\}\), \( \{\Pi_{1}, \Pi_{2},\dots, \Pi_{J}\}\subseteq \mathcal{P}\).

The proposed algorithm uses the carrier-to-interference (or \(C/I\)) values of the edge users to form priority and candidate lists to indicate the order in which possible coalitions will be tested. Essentially, a \(C/I\) value represents the macro-scale Signal-to-Interference Ratio (SIR) of a mobile user and is a metric of the intercell interference that it receives from the neighbouring RRHs. To examine the most beneficial coalitions, these values are compared to a threshold, i.e. the \(C/I\) threshold. If all the examined \(C/I\) values exceed this threshold, the proposed algorithm terminates, resulting in reduced number of steps of the proposed algorithm. The \(C/I\) value for UE$_{l,k_{e}}$ and an interfering RRH$_{l'}$ in dB is defined as:
\begin{equation}
C/I_{l',k_{e},n}=v_{l',k_{e},n}-v_{l,k_{e},n} ,
\end{equation}
where \(v_{l,k_{e},n}\) is calculated as in (4) and indicates that the \(C/I\) values are a function of macro-scale parameters such as path loss, shadowing and the transmitter and receiver antenna gains. To get a total metric of the intercell interference that all the edge users of an RRH receive from its neighbouring RRHs, if an RRH serves more than one edge user, the \(C/I\) values of the edge UEs of the same RRH are averaged at the cloud.

More specifically, the proposed coalition formation \textbf{Algorithm 1} consists of three main stages:

\begin{enumerate}

\item Initially, the small cell network consists of \(L\) non-cooperating RRHs, referred to as singleton coalitions.

\item At the second stage, every edge UE calculates an \((L-1) \times 1\) vector of \(C/I\) values between its serving and interfering RRHs and forwards them to the C-RAN. The interfering RRHs are assigned a unique
ID by each edge user and these IDs are also forwarded to the C-RAN.
If an RRH serves more than one edge UEs then, their \(C/I\) values are averaged, resulting into a total of \(L\) carrier-to-interference vectors, one for each RRH. Then, after these vectors are sorted in ascending order, they are combined into an \((L-1)\times L\) carrier-to-interference matrix \(\boldsymbol{I}_{M}\).

\item At the third stage, the first row of \(\boldsymbol{I}_{M}\) is extracted, its values are compared, and a priority list is formed, indicating the order in which each RRH will seek a collaborator, alongside with a candidate list, indicating potential collaborators. After that, the indicated coalitions are investigated sequentially by coordinating their members via the JT-CoMP technique.
A coalition is formed if the achieved payoff of at least one of the participant RRHs improves without hurting the payoff of the others, its utility is greater than the utilities of the previously formed coalitions of the examined collection and its size does not exceed the predefined maximum coalition size (merging functionality).

If a coalition has already been tested, then the next one (based on the priority list) is examined. This results in reduced steps for the proposed algorithm. Also, if an RRH is already a member of a non-singleton coalition and its turn, based on the priority list, comes, then the total coalition between the new candidate and all the existing members of the RRH's coalition is tested.
The same process repeats for every row of \(\boldsymbol{I}_{M}\) unless all \(C/I\) values exceed a predetermined \(C/I\) threshold.

Then, possible splits of the existing coalitions are examined. A split will occur only if by splitting a coalition at least one RRH will improve its payoff without decreasing the payoff of other members (splitting functionality).
\end{enumerate}

\begin{algorithm}[t!]
\caption{Coalition Formation Algorithm}
\begin{algorithmic}
\STATE 1: \textbf{Initial Stage}:
\bindent
\STATE The network consists of \(L\) singleton coalitions.
\eindent
\STATE 2: \textbf{Carrier-to-Interference Matrix Formation Stage}:
\bindent \FORALL{edge UEs} 
\STATE Calculate the \(C/I\) values from all the interferers.
\ENDFOR
\FORALL{\(l\in \mathcal{L}\)} 
\IF{attached edge UEs \(\geq\)  2} 
\STATE Average their edge UEs' \(C/I\) values.
\ENDIF
\ENDFOR
\STATE Form the \(C/I\) matrix \(\boldsymbol{I}_{M}\).
\eindent
\STATE 3: \textbf{Coalition Formation Stage}:
\bindent
\FOR{every row of \(\boldsymbol{I}_{M}\)}
\STATE Extract the row's \(C/I\) values.
\STATE Form the priority list \(\boldsymbol{p}\) and candidate list \(\boldsymbol{c}\).
\STATE Extract the examined pairs \(\boldsymbol{V}\) from \(\boldsymbol{p}\) and \(\boldsymbol{c}\).
\IF{any extracted \(C/I\) value \(\leq\) \(C/I_{threshold}\)}
\STATE (a)
\FOR{every \(l \in \boldsymbol{V}\)}
\STATE calculate \(\phi_{l}(\Pi_{i})\)
\IF{\(\phi_{l}(\Pi_{i})\) increases via Pareto order, \(u(\Pi_{i})\) increases and coalition size \(\leq\) max coalition size} 
\STATE Merge.
\ENDIF 
\ENDFOR
\ENDIF
\ENDFOR
\STATE (b)
\FOR{every \(\Pi_{i} \in \mathcal{P}\)}
\FOR{every \(l\in \Pi_{i}\)}
\STATE calculate \(\phi_{l}(\Pi_{j})\) after split, where \(\Pi_{j}\) a singleton coalition.
\IF{\(\phi_{l}(\Pi_{j})\) increases via Pareto order} 
\STATE Split.
\ENDIF
\ENDFOR
\ENDFOR

\eindent
\end{algorithmic}
\end{algorithm}

To adapt to network changes, such as handovers and user status changes due to mobility, two external rules were adopted to allow the coalitions to dynamically update their participant RRHs:
\begin{itemize}
\item  When a handover for a moving UE occurs or a UE changes its edge status, the proposed coalition formation algorithm is reactivated.

\item  When the algorithm is reactivated, instead of \(L\) singleton coalitions, the network consists of the coalitions formed from the previous activation of the proposed algorithm.

\end{itemize}

Compared to \cite{b51x,b52x} and \cite{b53x}, where each coalition was examined for 10 TTIs, in this study all the coalitions indicated by the proposed algorithm are examined instantaneously. This is possible as zero feedback delay is assumed and allows the study of the upper bound of the performance gain achieved by the proposed scheme. To adapt to this change, the splitting operations were moved and tried after all the rows of \(\boldsymbol{I}_{M}\) are examined for possible merges of RRHs, as in this case, the coalitions are not tested over time and the constant checking for possible splits is not needed. This means that the payoff of an RRH is a function of the instantaneous throughput values of its connected UEs. All the information about the coalition formation game, e.g. the record of attempted coalitions, the IDs of the RRHs inside a tested coalition, the throughput values for each UE of the involved RRHs before and after a coalition is tested, is kept in the proposed C-RAN functionality.

Next, we provide some definitions and properties of the proposed algorithm about its stability, convergence and complexity.

\textbf{Definition 9}: A partition \(\mathcal{P}\)  is $\mathbb{D}_{hp}$-stable if no players in \(\mathcal{P}\) are interested in leaving \(\mathcal{P}\) through merge and split to form other partitions in \(\mathcal{L}\) (not necessarily by merge and split) \cite{b54x}.

\textbf{Algorithm property 1}: Every partition resulting from the proposed
merge and split algorithm is $\mathbb{D}_{hp}$-stable.

\emph{Proof}: A network partition \(\mathcal{P}\) resulting from the proposed coalition formation algorithm can no longer be subject to any additional merge or split operations as successive iterations of these operations terminate \cite{b54x}. Therefore, the players in the final network partition \(\mathcal{P}\) cannot leave this partition through merge and split, hence the partition is $\mathbb{D}_{hp}$-stable.

\textbf{Algorithm property 2}: The coalition formation process with the proposed algorithm converges to a final stable partition in a finite number of steps.

\emph{Proof}: As the number of players is finite and a \(C/I\) threshold constrains algorithm's iterations, the number of steps of the proposed algorithm is finite. Additionally, any arbitrary sequence of the merge and split rules is guaranteed to converge to a final partition of \(\mathcal{L}\) \cite{b54x}.

\textbf{Algorithm property 3}: The complexity order of the proposed
coalition formation algorithm is \(\mathcal{O}(L_{n})\).

\emph{Proof}: The complexity order of the proposed algorithm depends on the number of merge and split operations, with each such operation being an iteration of the algorithm. The \(C/I\) threshold ensures a reduced number of iterations, as only the stronger interferers of the UEs of each RRH are taken into account. This means that exhaustive search is avoided. Also, the maximum coalition size ensures a reduced amount of possible splits. Thus, the complexity order of the proposed algorithm is \(\mathcal{O}(L_{n})\), where \(L_{n}\) is the average number of neighbouring RRHs that create significant interference to each RRH's users based on the \(C/I\) threshold.

\section{Other implementation issues}

To implement MU-MIMO and NOMA, beamforming, user pairing and power allocation need to be determined. The selection of the appropriate techniques is crucial, as the intrabeam and interbeam interference that the users receive can be significantly reduced by exploiting the CSI reported by them to the BBU pool. This can increase the users' per-subcarrier and per-OFDM symbol SINR and, ultimately, their throughput. In this section, we define how beamforming is done to mitigate interbeam interference, how the per-user power coefficients \(a_{l,n,k}\) are assigned and the method for the NOMA user pairing to minimize the intrabeam interference. Also, the selected multi-user scheduling method is presented.

\subsection{Beamforming}

To cancel the interbeam interference for users served by different beams of the same RRH, Zero-Forcing (ZF) beamforming can be deployed. ZF beamforming aims to minimize the interbeam interference inside a cell by generating beamforming vectors orthogonal to all the users' channel vectors except the corresponding cell's users \cite{b55x}.
Assuming perfect CSI, the number of transmit antennas should be larger than or equal to the number of total receive antennas in a cell, in order to have enough degrees of freedom on the transmitter side to generate a beamforming matrix that eliminates the interbeam interference \cite{b56x}, \cite{b57x}. In this paper, each RRH is equipped with \(N\) antennas, meaning that it can generate \(N\) beams. Each beam can serve at least two paired NOMA users scheduled in the same RB, which, in turn, means that this condition cannot be satisfied.

To implement ZF beamforming in NOMA, for each RB we select \(N\) scheduled users (one from each beam) with the higher channel gain. This selection is justified by the fact that their decoding order in their serving NOMA cluster is last and if they would receive interbeam inteference, the process for correctly decoding their received signal via SIC would be seriously affected \cite{b10x,b58x}. Then we calculate the beamforming vectors at each RB such as:

\begin{equation}
\boldsymbol{h}_{l,n,K}\boldsymbol{w}_{l,n'}= 0,\ n \neq n' .
\end{equation}

To satisfy (17), first, we define the matrix \(\boldsymbol{H}_{l}=[\boldsymbol{h}_{l,1,K}^{T}, \dots \boldsymbol{h}_{l,n,K}^{T}, \dots, \boldsymbol{h}_{l,N,K}^{T} ]^{T} \) and then calculate the Moore-Penrose pseudo-inverse matrix of \(\boldsymbol{H}_{l}\), i.e.:

\begin{equation}
(\boldsymbol{H}_{l})^{\dagger}=\boldsymbol{W}_{l}=(\boldsymbol{H}_{l})^{H} \Big((\boldsymbol{H}_{l})(\boldsymbol{H}_{l})^{H}\Big)^{-1} .
\end{equation}

Then, the beamforming vector \(\boldsymbol{w}_{l,n}\) for UE$_{l,n,K}$ can be obtained by normalizing the \(n\)-th column of \(\boldsymbol{W}_{l}\) \cite{b59x}.
Based on this choice of forming the beamforming vectors, only the user with the highest channel gain in each beam will avoid the interbeam interference. 
\subsection{Power coefficients allocation}

Power domain NOMA multiplexes users in the power domain. This means that the distribution of a beam's transmit power impacts significantly the performance of all the beam's users.
In our paper, for the non JT-CoMP case, each beam serves two NOMA users. When JT-CoMP is activated the number of users may vary. This means that a power allocation scheme which allows a fair extraction of the power coefficients is needed. However, the application of an optimal power allocation algorithm can increase the computational complexity so, a sub-optimal solution to this problem needs to be adapted. A suitable technique is the Fractional Transmit Power Control (FTPC) method \cite{b43x}, \cite{b46x}. Based on this technique, the power coefficient of a user \(k\), who is served by RRH \(l\), inside beam \(n\) for the subcarriers of an RB is allocated according to the following equation:
\begin{equation}
a_{l,n,k}=\frac{\displaystyle1}{\sum_{j=1}^K\Big({G_{l,n,j}}\Big)^{-p_{FTPC}}}\Big({G_{l,n,k}}\Big)^{-p_{FTPC}} ,
\end{equation}

\noindent where \(p_{FTPC}\) (\(0\leq p_{FTPC}\leq 1\)) is the decay factor. When \(p_{FTPC}\)=0, the power distribution for a beam's UEs is equal. The more the decay factor increases, the more power is allocated to the user with the lowest channel gain. This parameter can also be subject of optimization \cite{b43x}. Also, \(G_{l,n,k}\) for user \(k\), served by the \(l\)-th RRH in its \(n\)-th beam is defined as:
\begin{equation}
G_{l,n,k}=\|\boldsymbol{h}_{l,n,k}\|^2 .
\end{equation}

\subsection{NOMA user pairing}

Initially, before JT-CoMP is applied, we assume that each beam will serve exactly two users that are scheduled in the same RB, where one is considered the strong and one the weak user. This distinction is made based on the channel gains of the users, where the strong user's channel gain is greater than the weak user's. For a strong user, SIC and ZF beamforming will eliminate its received intrabeam and interbeam interferences respectively. Therefore, it is particularly important to define an efficient user pairing strategy in order to mitigate the intrabeam and interbeam interference for the weak UEs as well. 

Based on equations (6), (10), (12) and (19), 
we can easily see that the greater the difference between the channel gains of the strong and the weak user, the more the performance improvement for a weak user. This is a consequence of the decreased power coefficients for the strong user, resulting in reduced intrabeam interference for the weak one. Also, based on \cite{b58x}, when ZF beamforming is applied and the channel \(\boldsymbol{h}_{l,n,k_{s}}\) of the strong user is highly correlated with the channel \(\boldsymbol{h}_{l,n,k_{w}}\) of the weak user, i.e. \(\boldsymbol{h}_{l,n,k_{s}} \approx c\cdot \boldsymbol{h}_{l,n,k_{w}}\), where \(c\) is a constant, then \(\boldsymbol{h}_{l,n,k_{w}}\boldsymbol{w}_{l,n'} \approx 0\) for \(n \neq n'\), which means that the interbeam interference for the weak user can be significantly mitigated when this NOMA pair is formed.
Thus, the NOMA user pairing for the subcarriers and OFDM symbols of each RB is done in two stages. At the first stage the scheduled users in the RB are divided into strong users and weak users based on their channel gain. Then, in the second stage, each strong user forms a NOMA pair with the weak user that maximizes its correlation metric. The correlation metric is given via the following equation \cite{b58x}:
\begin{equation}
C_{k_{s},k_{w}}=\frac{\|\boldsymbol{h}_{l,n,k_{s}}\boldsymbol{h}_{l,n,k_{w}}^{H}\|}{\|\boldsymbol{h}_{l,n,k_{s}}\|\|\boldsymbol{h}_{l,n,k_{w}}\|} .
\end{equation}

When JT-CoMP is activated, depending on the beam assignment, each beam can serve more than two users, as edge UEs will be served by all the RRHs in a coalition for their scheduled RBs. Again, for JT-CoMP, the edge UEs are assigned to the NOMA clusters of the cooperating RRHs that maximize the correlation metric given by (21) in order to mitigate the interference from other beams of the cooperating RRHs, except the serving ones.

It is important to note that the distinctions between an edge and non-edge user and strong and weak user are entirely different, e.g. based on its effective SINR value, a strong user may be an edge user and a weak user may be a non-edge user.

\subsection{Multi-user scheduling}

In NOMA, the scheduler pairs multiple users for simultaneous transmission at each RB. To determine the set of users that are going to be served in the same RB, a Round Robin scheduler was implemented. At each TTI, the BBU pool creates a random sequence of unique UE IDs for each RRH which includes all the UEs connected to it. Then, according to the number of users per NOMA cluster and transmit antennas, a group of \(K \cdot N\) UEs is selected to be served per RB in each cell. After that, the scheduled users in each cell are organized in \(N\) NOMA clusters according to the selected user-pairing method. For the following RB, the next \(K \cdot N\) users are selected and the same procedure repeats by scheduling groups of UEs in a cyclic manner until the scheduler allocates all the available RBs at each cell.

\section{Experimental performance analysis}
In this section, the simulation results are shown and analyzed, after the simulation setup and the examined scenarios are presented.
\subsection{Simulation setup}
A small cell C-RAN network is assumed, where each RRH's total transmit power was set at 30 dBm and their coverage was assumed omnidirectional. Each RRH was equipped with 4 transmit antennas, and thus can generate 4 beams which serve initially, i.e. before JT-CoMP and the proposed coalition formation algorithm application, two NOMA users each in every RB. To distribute the available transmit power of each beam to the users of the beam at an RB, a decay factor of 0.4 was selected. Each RRH has an available bandwidth \(B\) of 20 MHz and a subcarrier spacing of 15 KHz was selected, corresponding to 106 RBs per TTI available for allocation to its connected users with a subframe and slot duration of 14 OFDM symbols for a normal cyclic prefix, i.e. 1 ms \cite{b42x}, \cite{b60x}, which is also the TTI duration. It is assumed that all the OFDM symbols per slot are allocated for the PDSCH. Also, an RB contains 12 subcarriers in the frequency domain, yielding an RB bandwidth of 180 KHz. The CQI indices and their interpretation regarding the MCS were based on the Table 5.2.2.1-3 of \cite{b48x}, meaning that a modulation scheme up to 256QAM can be selected. For the TB size calculation it is assumed that each slot contains two DM-RS symbols\cite{b61x}, there is no overhead and the number of MIMO layers is one. A carrier frequency of 3.5 GHz was assumed and each RRH's circumradius was set at 125 meters. Each RRH is assumed to initially serve 15 UEs, randomly placed in its cell area with positions based on a uniform distribution. The UEs are set to move in random directions, constant for the duration of each simulation, with a speed of 5 km/h. For the simulation scenario, the WINNER II Urban Microcell (UMI) case path loss model was used with a shadowing standard deviation of 4 dB \cite{b62x}.

For the proposed algorithm's settings, a maximum coalition size of 4 was assumed to guarantee small feedback overhead and backhaul traffic \cite{b14x}, while the \(C/I\) threshold was set at 10 dB. The \(C/I\) threshold was set at a high value to account for the randomness of the UE positions. For example, an edge user, based on its position in the cell, might receive high interference from an RRH, resulting in a low individual \(C/I\) value, while another user might receive low interference, resulting in a high individual \(C/I\) value. The cloud averages the edge UEs' \(C/I\) values to represent the interference that all the edge UEs of an RRH receive, so in the case that the averaged \(C/I\) value is high and the \(C/I\) threshold is low, a coalition between the serving RRH and an RRH that creates severe interference for an edge UE might not be examined. However, testing more coalitions means more iterations, which is why this parameter was not set at a higher value.

\begin{figure}[t!]
\centerline{\includegraphics[width=1\columnwidth,height=\textheight,keepaspectratio]{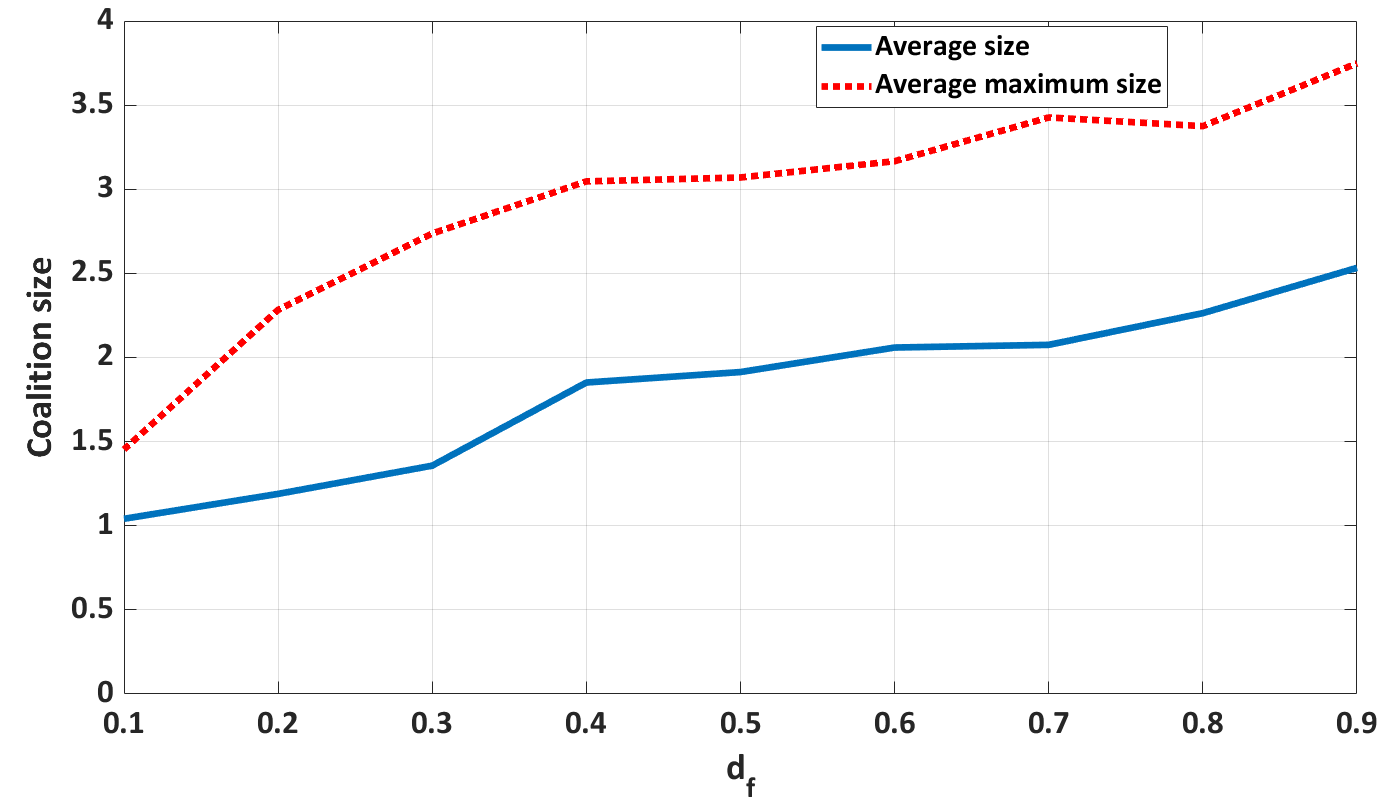}}
\caption{Average and average maximum coalition sizes versus the acceptable non-edge UE throughput decrease when the proposed coalition formation game is activated.}
\label{fig3}
\end{figure}

\begin{figure}[t!]
\centerline{\includegraphics[width=1\columnwidth,height=\textheight,keepaspectratio]{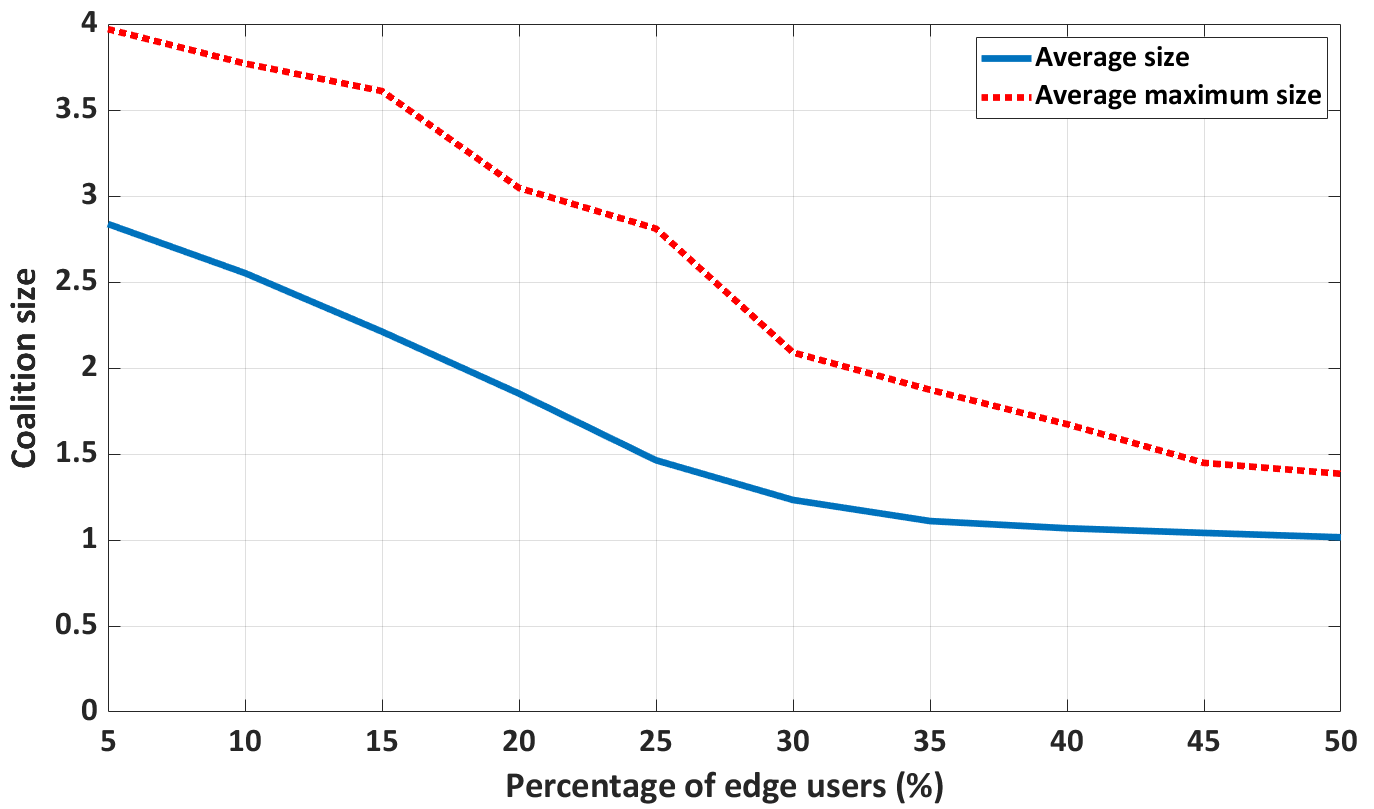}}
\caption{Average and average maximum coalition sizes versus the
percentage of edge UEs in the network when the proposed coalition formation game is activated.}
\label{fig4}
\end{figure}

The most important parameter for the proposed coalition formation algorithm is \(d_{f}\) which represents the acceptable throughput decrease for the non-edge UEs of a formed coalition. To determine an appropriate value for this parameter, various simulations for different values of \(d_{f}\) were run. Specifically, we run 100 simulations of a 12 RRH scenario with no user mobility for each value of \(d_{f}\) between 0.1 and 0.9, with a step of 0.1. Each simulation had a duration of 100 TTIs. The results in terms of the average and maximum coalition size are depicted in Fig.~\ref{fig3}. Based on Fig.~\ref{fig3}, the formation of coalitions for values of \(d_{f}\) less than or equal to 0.3 is very difficult which is reflected in the average coalition size which is below 1.5. Therefore, we chose the lowest value that guarantees a balance of minimum non-edge UE throughput decrease and a satisfying average and maximum coalition size. This value is \(d_{f}= 0.4\) which means that when the proposed coalition formation game is activated, a coalition between RRHs can be formed when it does not result in a throughput reduction more than 40\% for any non-edge UE served by the coalition's RRHs, compared to its throughput without the JT-CoMP activation.

Instead of setting a predetermined SINR threshold to distinguish the edge from the non-edge UEs, this value is determined based on the highest effective SINR value between a percentage of UEs with the lowest effective SINR values before the algorithm is run and JT-CoMP is applied. This prohibits the dense population of the network by edge UEs which can make the formation of coalitions with the proposed algorithm much more difficult, as the available resources, i.e. the transmit power and the RBs, are limited \cite{b51x,b52x} and their distribution among a large number of users can undermine the UE performance. Also, not having a predetermined SINR threshold can account for the seldom cases that most users are located near the edge or the center of their cell where this would result in too many or hardly any edge UEs respectively. To determine the appropriate value for the edge UE percentage in the network, we run simulations where the proposed algorithm is applied for edge UE percentages between 5\% and 50\% with a 5\% step. For each percentage value, 100 simulations with a duration of 100 TTIs each were run where \(d_{f}\) was set at 0.4, 12 RRHs were assumed in the network and the users were considered immobile. Fig.~\ref{fig4} depicts the average and average maximum coalition size for various percentage values of edge users in the network. Based on Fig.~\ref{fig4}, this percentage was set at 20\%, as we seek a moderate edge UE load in order to help the proposed algorithm to form as many coalitions as possible, while helping as many users as possible to increase their throughput. Furthermore, this value was used to extract Fig.~\ref{fig3} and determine the acceptable throughput decrease value for the non-edge UEs, i.e. \(d_{f}\). A summary of the simulation parameters is presented in Table~\ref{tab1}. 

\begin{table}[htbp]
\caption{Simulation Parameters}
\label{Table I}
\begin{center}
\begin{tabular}{l l}
\hline
\textbf{Parameter description}&{\textbf{Value}} \\
\hline
Carrier frequency& 3.5 GHz\\
Transmit power& 30 dBm\\
Cell layout& Hexagonal grid (wrap-around)\\
Cell circumradius& 125 m\\
Transmitter height& 10 meters\\
Receiver height& 1.5 meters\\
Antenna gain pattern& Omni-directional\\
Number of Tx antennas& 4\\
Number of Rx antennas& 1\\
Total Tx antenna gain& 8.17 dBi\\
Total Rx antenna gain& 0 dBi\\
Path loss model& WINNER II UMI\\
Shadow fading standard deviation& 4 dB\\
Receiver type& SIC\\
Bandwidth& 20 MHz\\
Subcarrier spacing& 15 KHz\\
Available Resource Blocks& 106\\
No. of NOMA clusters per cell& 4\\
No. of UEs per cell& 15\\
Receiver noise figure& 9 dB\\
Thermal noise density& -174 dBm/Hz\\
\hline
\end{tabular}
\label{tab1}
\end{center}
\end{table}

Four simulation scenarios were examined and compared with each other in terms of user throughput. More specifically, the case where the proposed scheme with JT-CoMP and the proposed coalition formation algorithm, referred to as game JT-CoMP, is compared to a JT-CoMP scenario with predetermined and fixed (static) coalitions, referred to as SC JT-CoMP, a JT-CoMP scenario where a greedy clustering algorithm is implemented, referred to as GC JT-CoMP, and a scenario where JT-CoMP is not implemented in the system, referred to as the no JT-CoMP case. For the SC JT-CoMP case, a predetermined coalition size was selected and the only criterion for the coalition formation is the distance between the RRHs. If the number of RRHs in the network did not allow the formation of coalitions with the predetermined size, RRHs in the network edges were considered for these coalitions as their UEs receive the least intercell interference.

As for the GC JT-CoMP scenario, the proposed greedy clustering algorithm of \cite{b39x} was implemented. More specifically, starting from a random RRH, the greedy clustering algorithm examines all the possible coalition combinations that include it and don't exceed a predetermined maximum size and forms the ones that maximize the sum-throughput of the served edge UEs of the involved RRHs. Compared to \cite{b39x} we chose to maximize the sum-throughput of the edge UEs instead of the sum-throughput of all UEs. This is a conscious choice to examine whether the proposed scheme provides the most consistent improvements in edge UE throughput as, in the GC JT-CoMP case, possible individual throughput decreases for the edge UEs due to resource scarcity are not considered. As the greedy clustering schemes are considered dynamic \cite{b22x}, the greedy clustering algorithm is reactivated based on the first reactivation rule of the proposed coalition formation algorithm, meaning that the clustering process always starts considering singleton coalitions. For the SC JT-CoMP and GC JT-CoMP cases a maximum coalition size of 4 was selected, as it is the same for the proposed game JT-CoMP scheme. Simulations for each examined scenario were run 100 times for increased accuracy and each simulation had a duration of 1000 TTIs.

\subsection{Results}

\begin{figure}[t!]
\centerline{\includegraphics[width=1\columnwidth,height=\textheight,keepaspectratio]{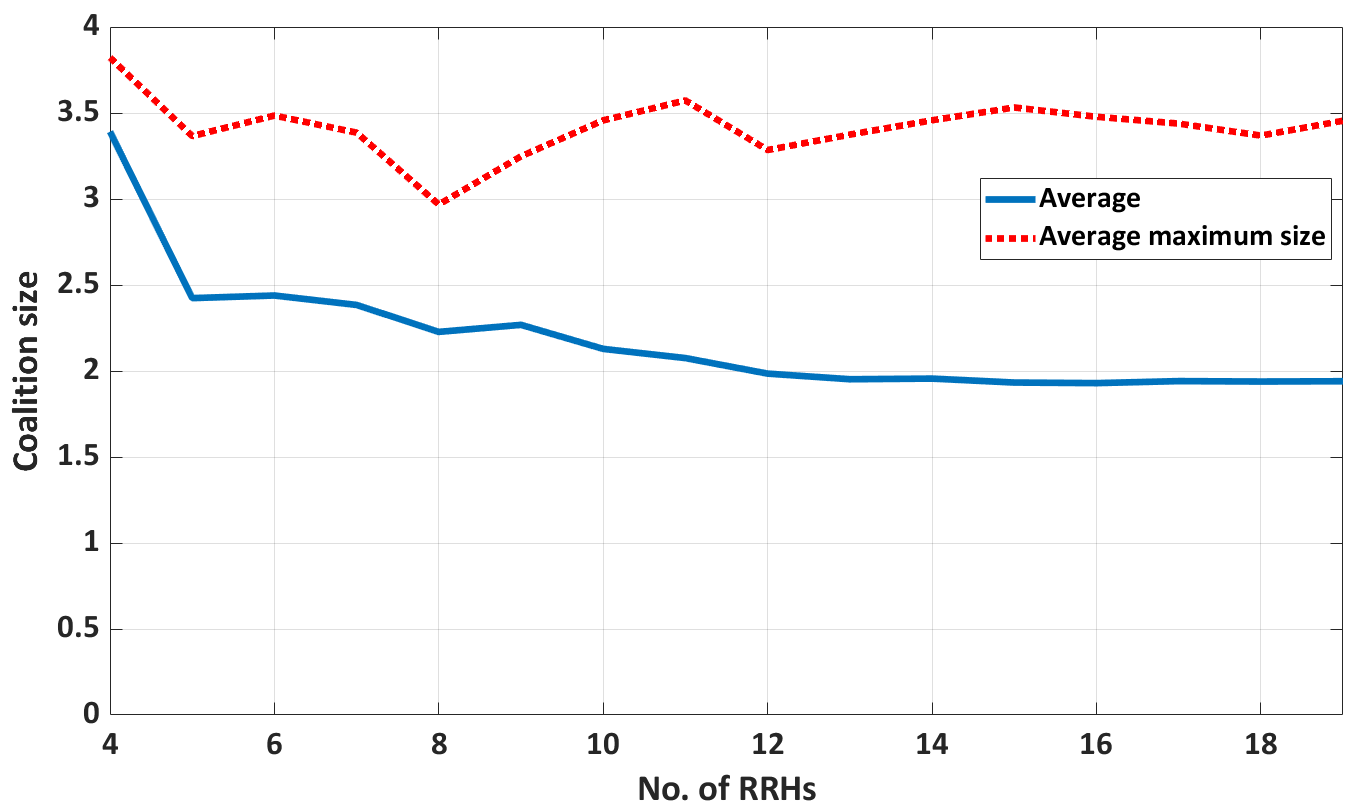}}
\caption{Average and average maximum coalition size versus the number of RRHs in the network when the proposed coalition formation game is applied.}
\label{fig5}
\end{figure}

Fig.~\ref{fig5} depicts the average and average maximum coalition size that occurs from the application of the proposed algorithm when the number of RRHs in the network varies. The proposed algorithm, combined with the appropriate settings, results in a relatively stable average coalition size, regardless of the number of RRHs in the network, while the maximum coalition size fluctuates slightly around 3.4. This certifies the ability of the proposed algorithm to form coalitions which result in increased edge user throughput and moderately decreased non-edge throughput. In fact, the average and average maximum coalition size was not much smaller than what the exhaustive search with the same conditions as the proposed algorithm yielded. Specifically, the exhaustive search yielded an average of just 1.3182\% and 1.6425\% higher average and average maximum coalition size respectively. This means that the selected \(C/I\) threshold allowed the examination and the formation of almost all the beneficial coalitions for the throughput of the edge and non-edge UEs of the network.

\begin{figure}[t!]
\centerline{\includegraphics[width=1\columnwidth,height=\textheight,keepaspectratio]{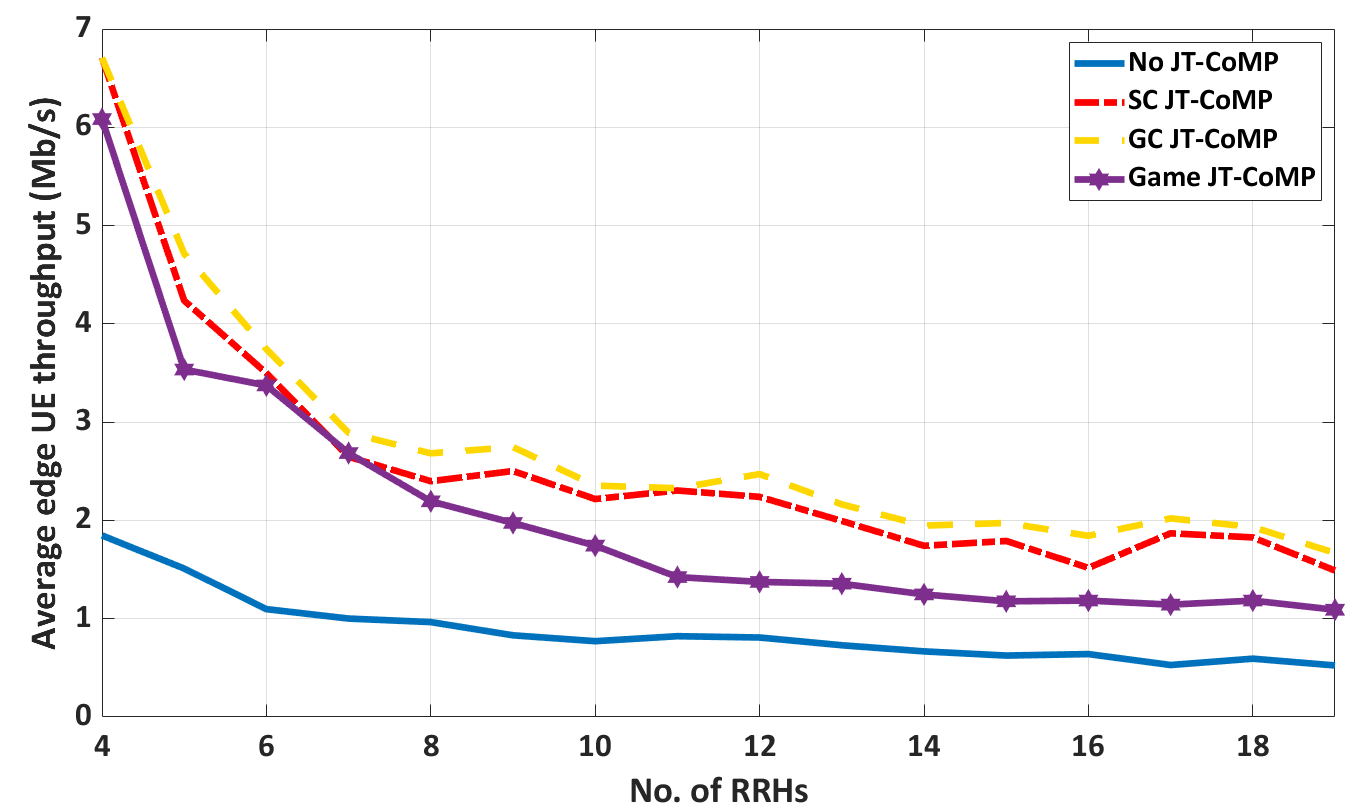}}
\caption{Average edge user throughput versus the number of RRHs in the network for all the examined cases.}
\label{fig6}
\end{figure}

\begin{figure}[t!]
\centerline{\includegraphics[width=1\columnwidth,height=\textheight,keepaspectratio]{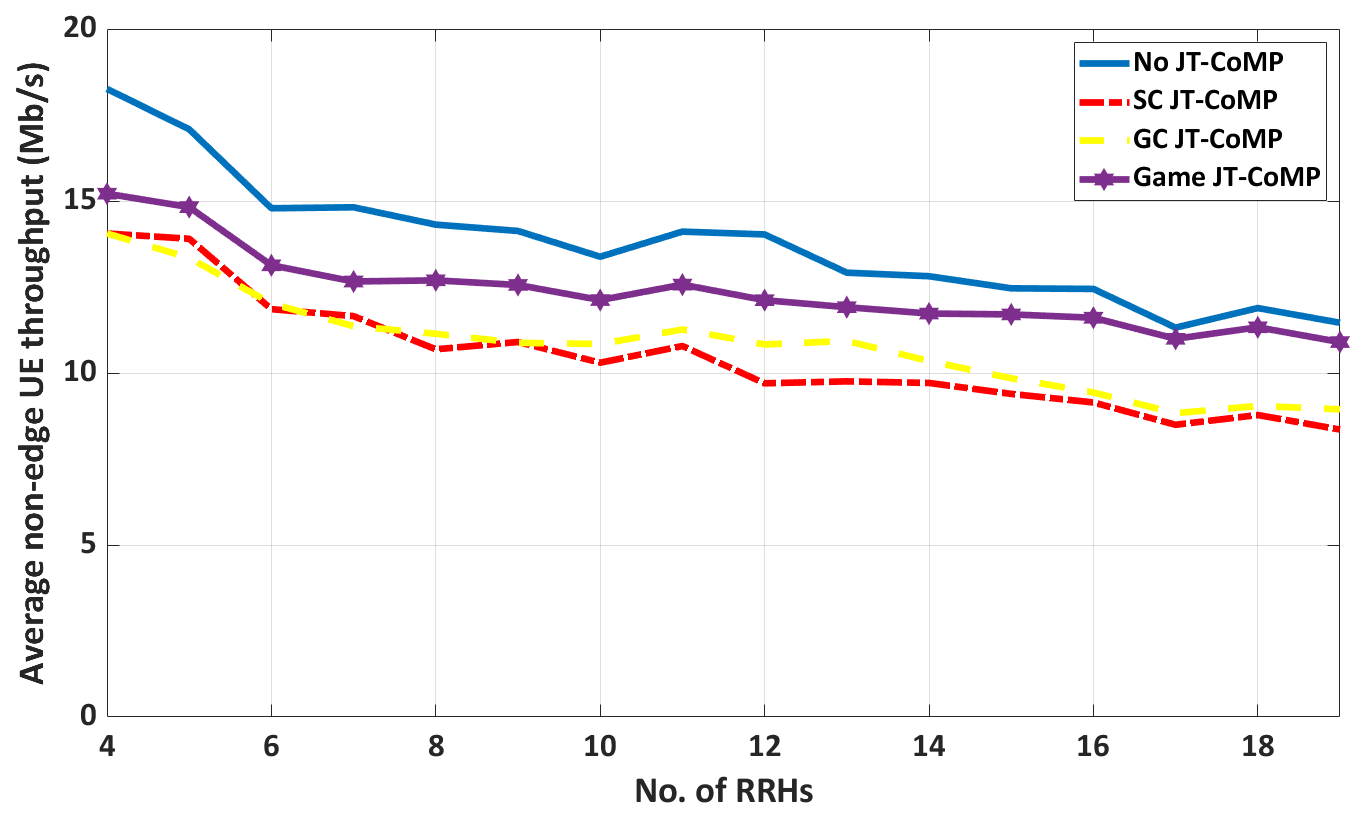}}
\caption{Average non-edge user throughput versus the number of RRHs in the network for all the examined cases.}
\label{fig7}
\end{figure}

Fig.~\ref{fig6} depicts the average edge user throughput for all the examined cases when the number of RRHs in the network varies. The guaranteed big coalition size of the SC JT-CoMP and GC JT-CoMP cases results in higher increase of the instantaneous throughput of several edge users when these schemes are applied compared to the proposed scheme of game JT-CoMP. However, the proposed scheme performs close in terms of average edge UE throughput compared to the SC JT-CoMP and GC JT-CoMP cases, even if the average and maximum coalition size in this case is less, as shown in Fig.~\ref{fig5}. This is a result of the incautious transmission point clustering of these schemes which may result in throughput reductions for some edge UEs. Compared to the no JT-CoMP case, all the JT-CoMP cases show vast improvement in terms of throughput as a logical result of the reduced intercell interference that the edge UEs receive. Between the GC JT-CoMP and SC JT-CoMP, the application of a greedy clustering algorithm guarantees a level of intelligence in the transmission point clustering compared to the static clustering method which is based only on the distance between RRHs, resulting in higher average edge UE throughput values for the GC JT-CoMP scheme. For all the examined cases, the increase in RRHs in the network results in more intercell interference, which justifies the decreasing average edge user throughput with an increasing number of RRHs in the network.

On the other hand, the game JT-CoMP case presents the best average non-edge user throughput results compared to the other JT-CoMP schemes, as it can be seen in Fig.~\ref{fig7}. This is expected, as the proposed coalition formation algorithm restricts how much the non-edge user throughput can be reduced. It is particularly interesting that in many cases, the GC JT-CoMP scheme outperforms the SC JT-CoMP scheme, despite also presenting better edge user throughput, certifying that forming fixed clusters of transmission points based only on distance in a dynamic system is far from an optimal solution. As for the edge users, more RRHs in the network means more intercell interference for the non-edge UEs as well, which causes the downturn in average non-edge UE throughput in Fig.~\ref{fig7}.

\begin{figure}[t!]
\centerline{\includegraphics[width=1\columnwidth,height=\textheight,keepaspectratio]{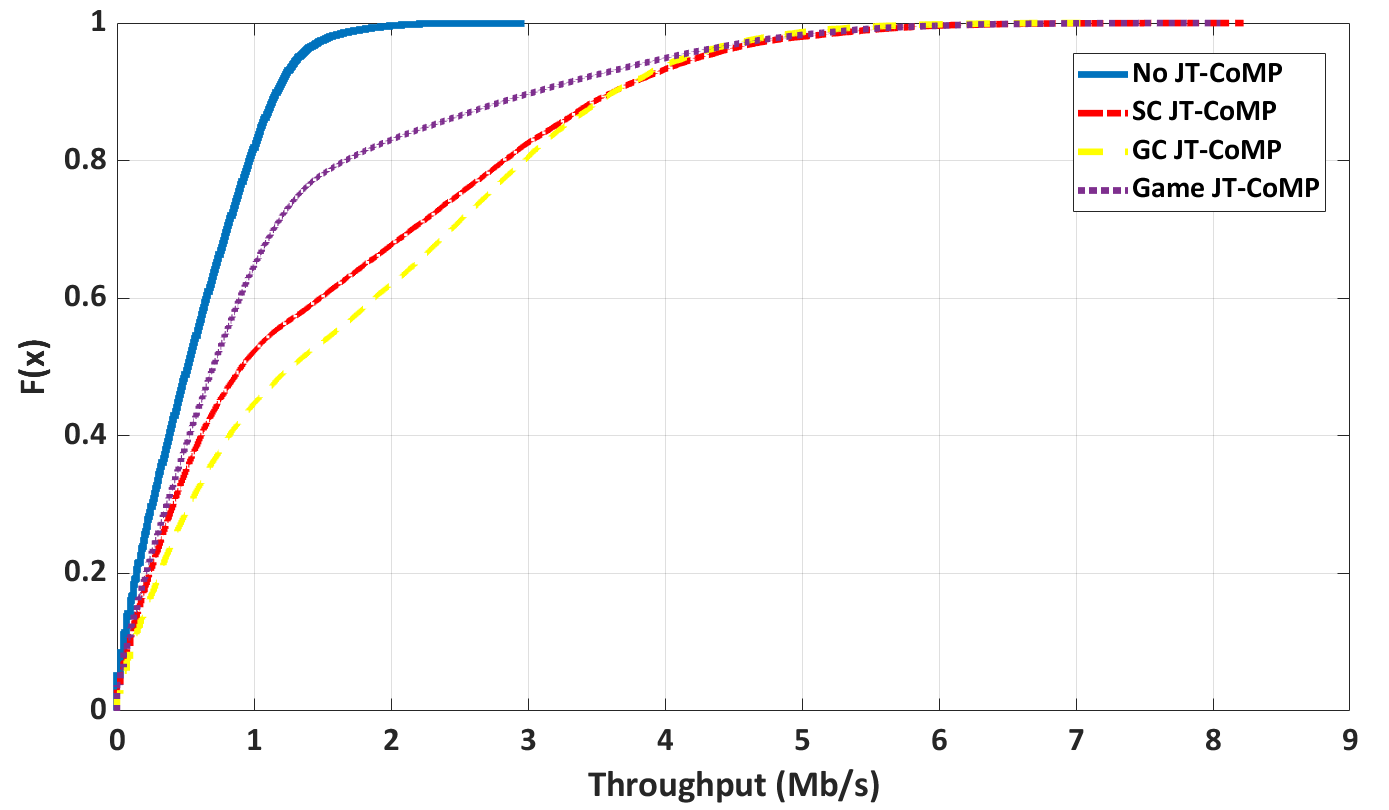}}
\caption{CDF of the individual instantaneous edge user throughput.}
\label{fig8}
\end{figure}

\begin{figure}[t!]
\centerline{\includegraphics[width=1\columnwidth,height=\textheight,keepaspectratio]{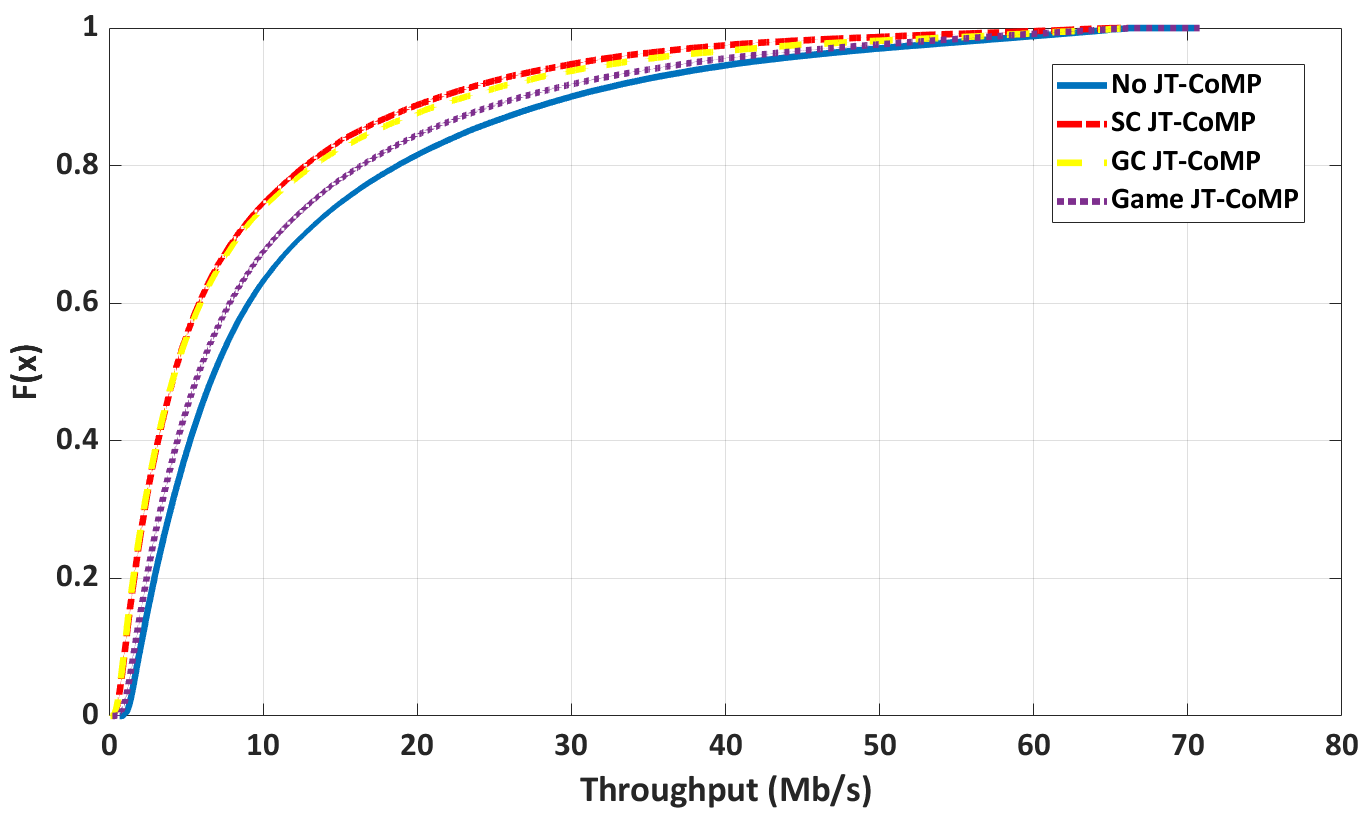}}
\caption{CDF of the individual instantaneous non-edge user throughput.}
\label{fig9}
\end{figure}

\begin{figure}[t!]
\centerline{\includegraphics[width=1\columnwidth,height=\textheight,keepaspectratio]{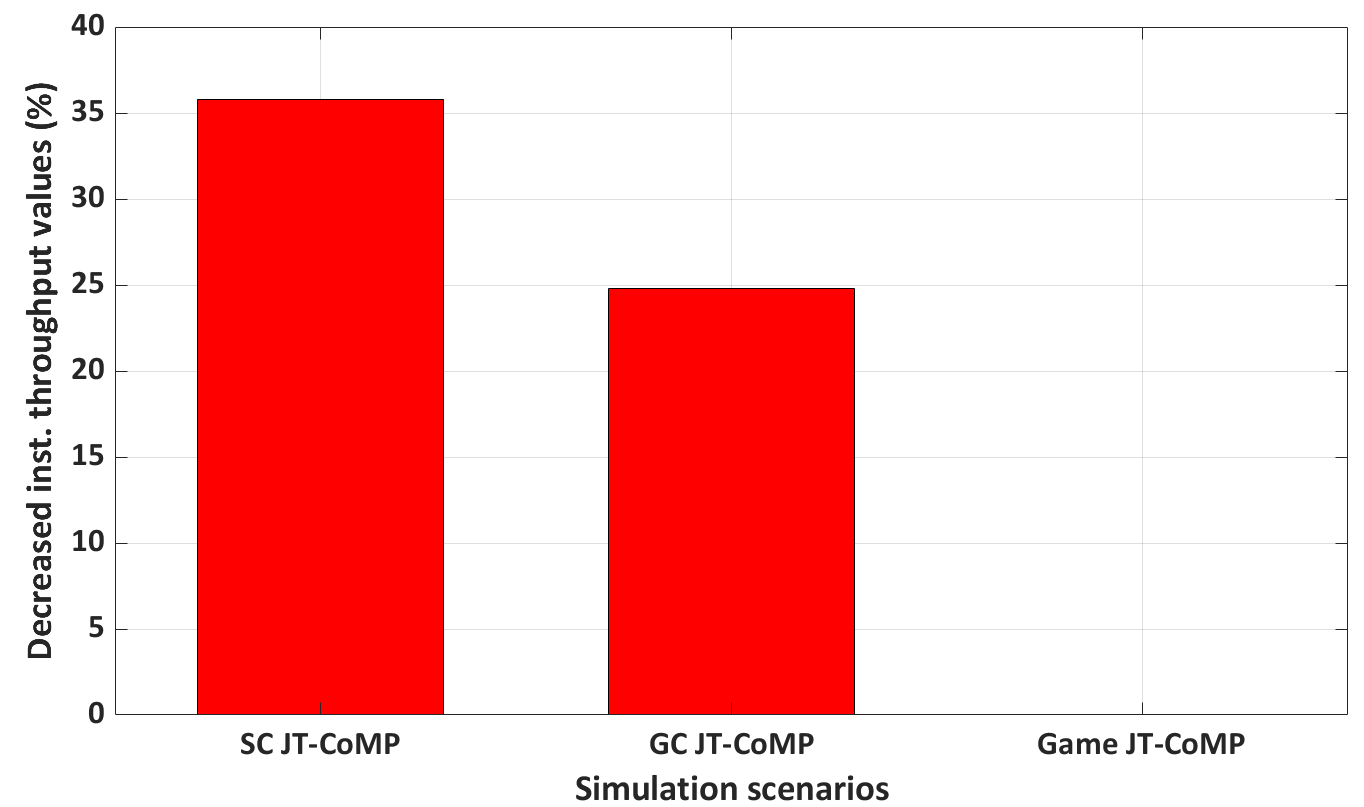}}
\caption{Percentage of decreased instantaneous throughput values compared to the no JT-CoMP case for all JT-CoMP cases.}
\label{fig10}
\end{figure}

\begin{figure}[t!]
\centerline{\includegraphics[width=1\columnwidth,height=\textheight,keepaspectratio]{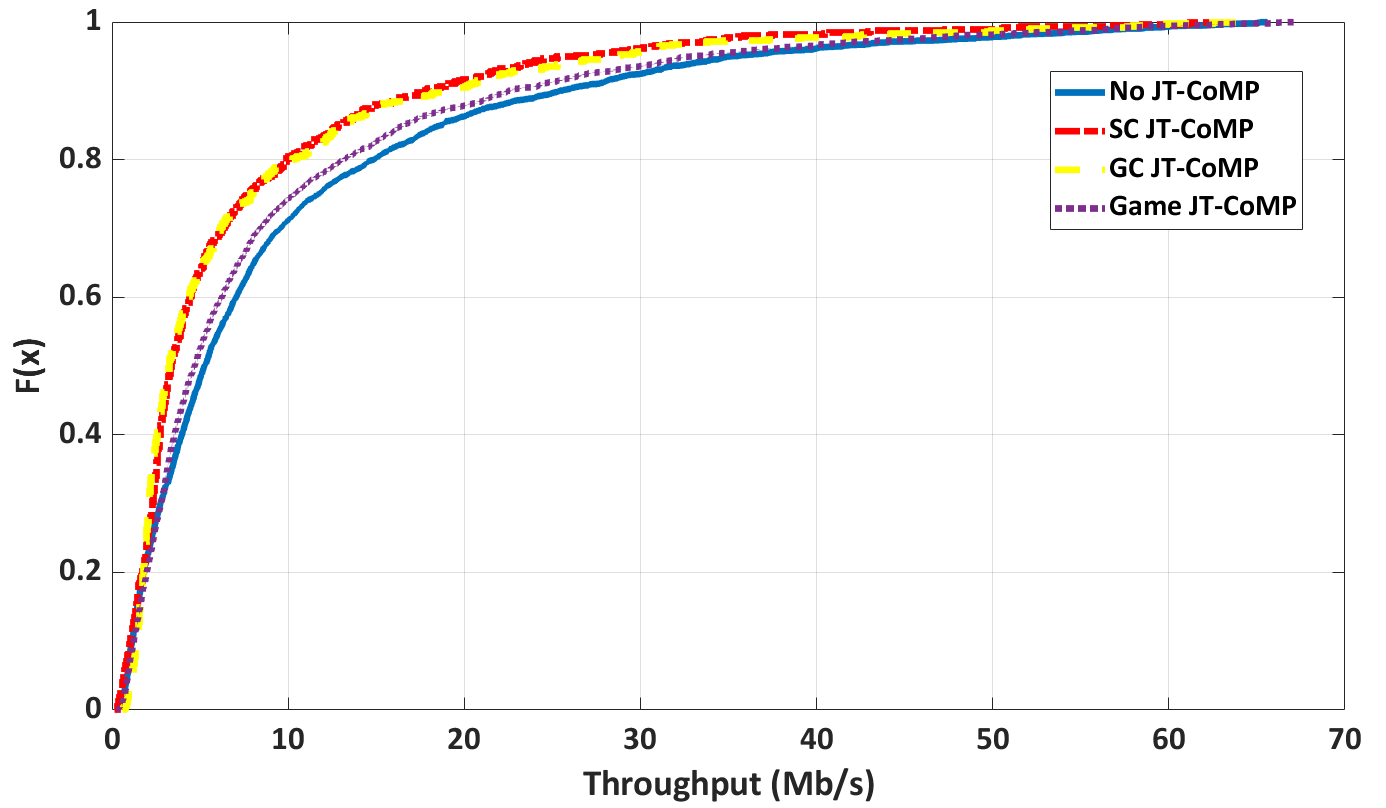}}
\caption{CDF of the  individual average throughput of all users.}
\label{fig11}
\end{figure}

The same conclusions can be extracted from Fig.~\ref{fig8} and Fig.~\ref{fig9}, where the Cumulative Distribution Function (CDF) of the individual instantaneous user throughput is depicted when the edge and non-edge users are considered respectively for a 19 RRH scenario. Furthermore, as shown by Fig.~\ref{fig10}, for the SC JT-CoMP and GC JT-CoMP cases, there was a significant percentage of individual instantaneous throughput values which were lower than the corresponding values if the JT-CoMP technique was not applied. This means that in these cases, even for a 20\% edge UE load in the network, the size of some NOMA clusters with JT-CoMP causes the distribution of the per-NOMA cluster allocated power to enough users to result in throughput reduction for some of them. Moreover, another factor that contributed to this throughput reduction is the intrabeam interference which increases for the previously existing UEs in the NOMA clusters when new edge users are added to them.

Fig.~\ref{fig11} depicts the CDF of the individual average throughput of all users in the network for a 19 RRH scenario. An improvement regarding average throughput can be observed for the low throughput users for the GC JT-CoMP and game JT-CoMP cases compared to the no JT-CoMP case. This is expected as, for the no JT-CoMP case, these users spent most of the duration of the simulations being edge UEs, which resulted in an increase of their throughput via the JT-CoMP technique. Moreover, it is noteworthy that the proposed scheme was able to provide throughput improvements for a wider range of low throughput cases compared to the GC JT-CoMP scheme, while having smaller feedback overhead and coordination costs, due to coalition size, and guaranteeing that the throughput of the non-edge UEs is not severely reduced. Regarding the SC JT-CoMP case, its incautious transmission point clustering resulted in lower individual average throughput values compared to the no JT-CoMP case, proving that intelligence needs to be behind the formation of transmission point coalitions.

\begin{figure}[t!]
\centerline{\includegraphics[width=1\columnwidth,height=\textheight,keepaspectratio]{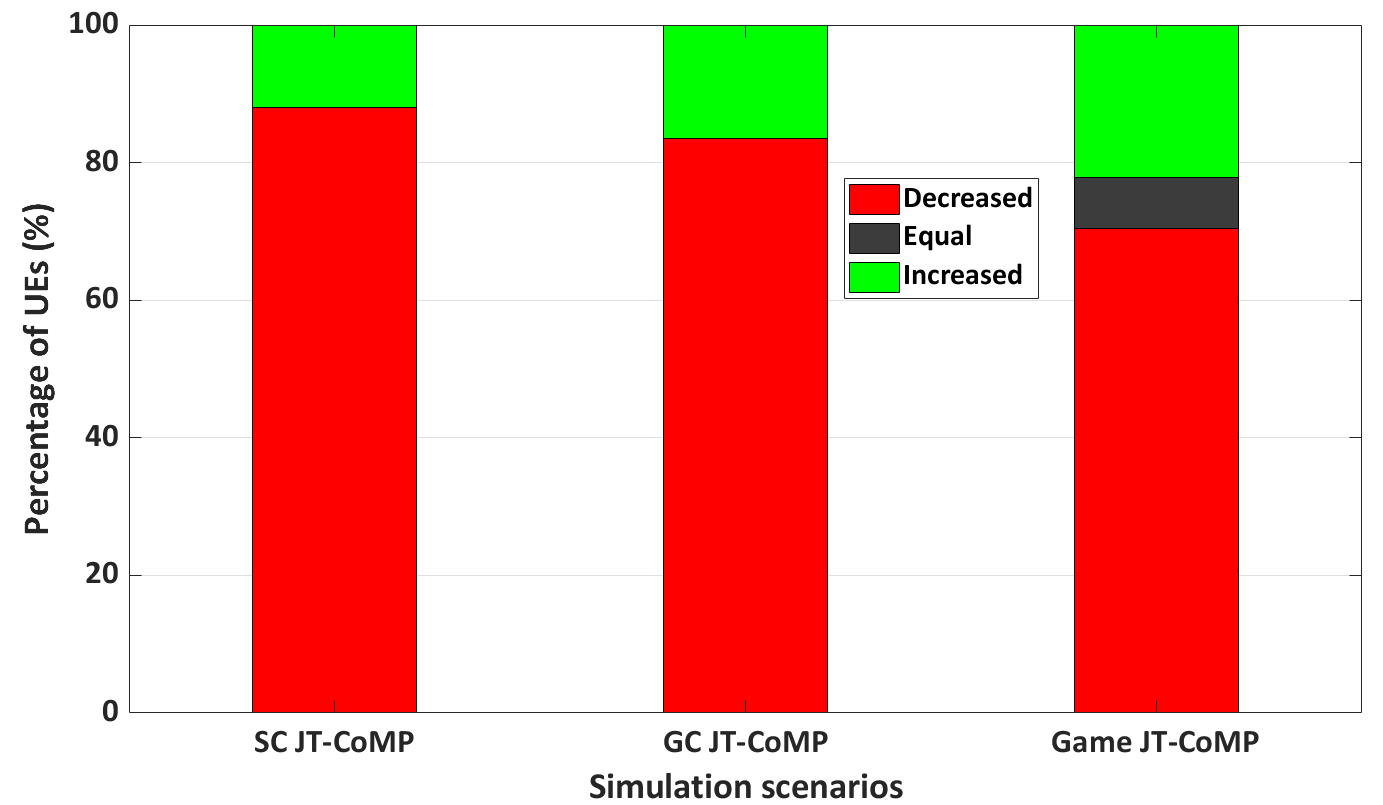}}
\caption{Percentage of users with decreased, equal and increased individual average throughput compared to the no JT-CoMP case for all JT-CoMP cases.}
\label{fig12}
\end{figure}

\begin{figure}[t!]
\centerline{\includegraphics[width=1\columnwidth,height=\textheight,keepaspectratio]{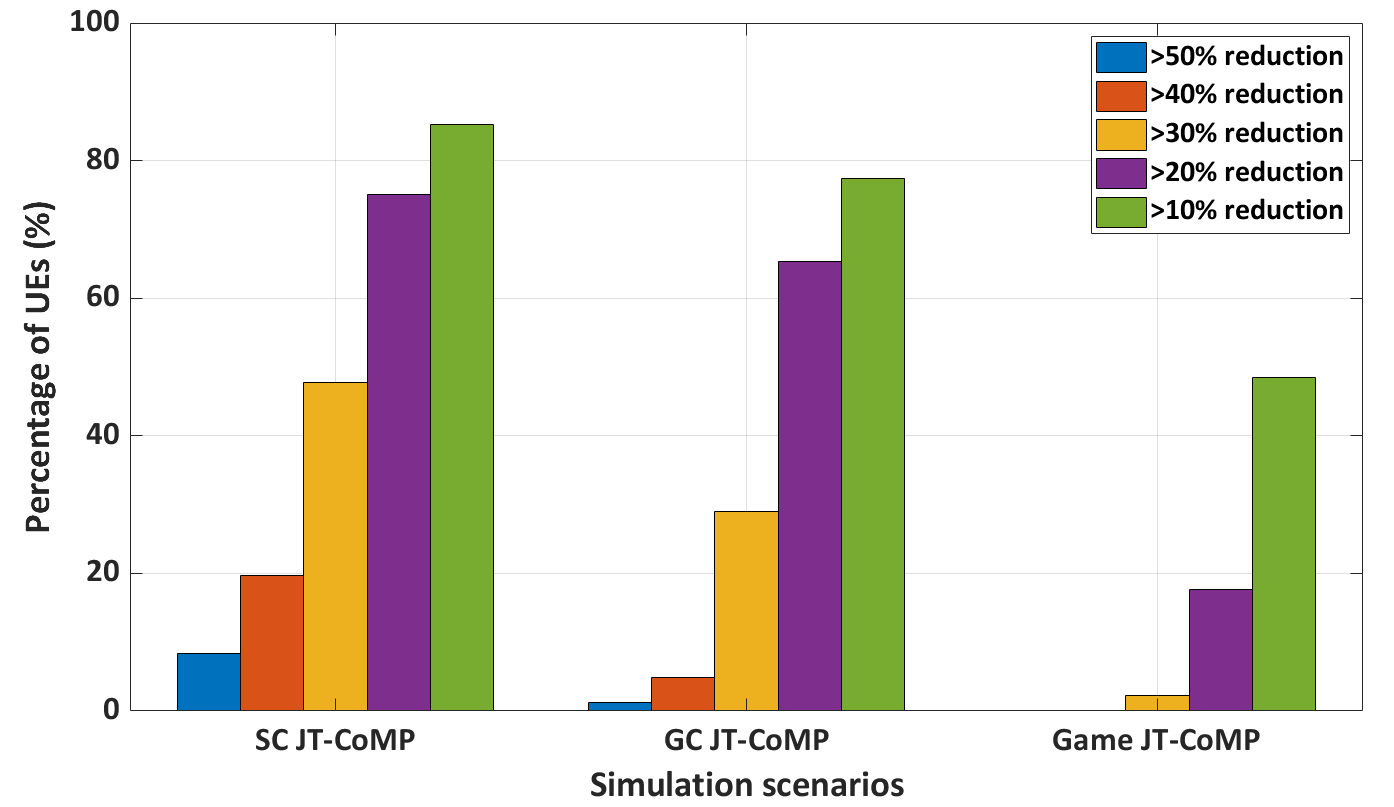}}
\caption{Percentage of users with individual average throughput reduced more than 50, 40, 30, 20 and 10\% compared to the no JT-CoMP case for all JT-CoMP cases.}
\label{fig13}
\end{figure}

Fig.~\ref{fig12} presents the percentage of the users with decreased, equal and increased individual average throughput compared to the no JT-CoMP case for all JT-CoMP cases when 19 RRHs were assumed in the network. The proposed scheme resulted in an increase in the average throughput of more users compared to the other JT-CoMP cases. Furthermore, the existence of singleton coalitions in the game JT-CoMP case, when the conditions of the proposed coalition formation algorithm are not satisfied, resulted in a small percentage of users having the same average throughput as in the case of no JT-CoMP.  Also, the fact that the status of any user can change during the simulations resulted in the individual average throughput increase, or not decrease, for more users than the 20\% that are considered edge per TTI. In contrast, the SC JT-CoMP and GC JT-CoMP cases have a fixed coalition size, which forced users to show either increased or decreased individual average throughput compared to the no JT-CoMP case. The fixed coalition size of these schemes also resulted in the formation of some overcrowded NOMA clusters, which in turn resulted in more UEs with reduced individual average throughput compared to the game JT-CoMP case, even if some of them were considered edge users for the majority of the duration of the simulations.

Regarding the individual average throughput reductions, Fig.~\ref{fig13} depicts the percentage of the users that had their average throughput decrease more than 50, 40, 30, 20 and 10\% for each JT-CoMP case compared to the no JT-CoMP scenario. Again, the network consisted of 19 RRHs. No user had its average throughput decreased more than 40\% for the game JT-CoMP case, which was expected, even if the conditions of the proposed algorithm pertain to the instantaneous throughput. Additionally, even the percentage of users with average throughput reduction greater than 30\% was almost null. On the other hand, a significant number of UEs had their average throughput reduce more than 40 and 30\% when the SC JT-CoMP and GC JT-CoMP schemes were applied. In a dynamic system, the status of the users can change constantly, meaning that a user can be considered edge and non-edge between two TTIs. By applying the proposed coalition formation algorithm, we guarantee that when the user is considered non-edge, its instantaneous throughput will not be severely reduced, whereas if the user is considered edge, its instantaneous throughput will be increased, or at least not decreased. In the other examined JT-CoMP schemes, the instantaneous throughput of a user may increase more when the UE is considered edge and the NOMA clusters that it is assigned are not overcrowded. However, when the user is assigned in overcrowded NOMA clusters or becomes non-edge, its instantaneous throughput can plummet. This results in higher individual average throughput reductions, as shown in Fig.~\ref{fig13}. 

\begin{figure}[t!]
\centerline{\includegraphics[width=1\columnwidth,height=\textheight,keepaspectratio]{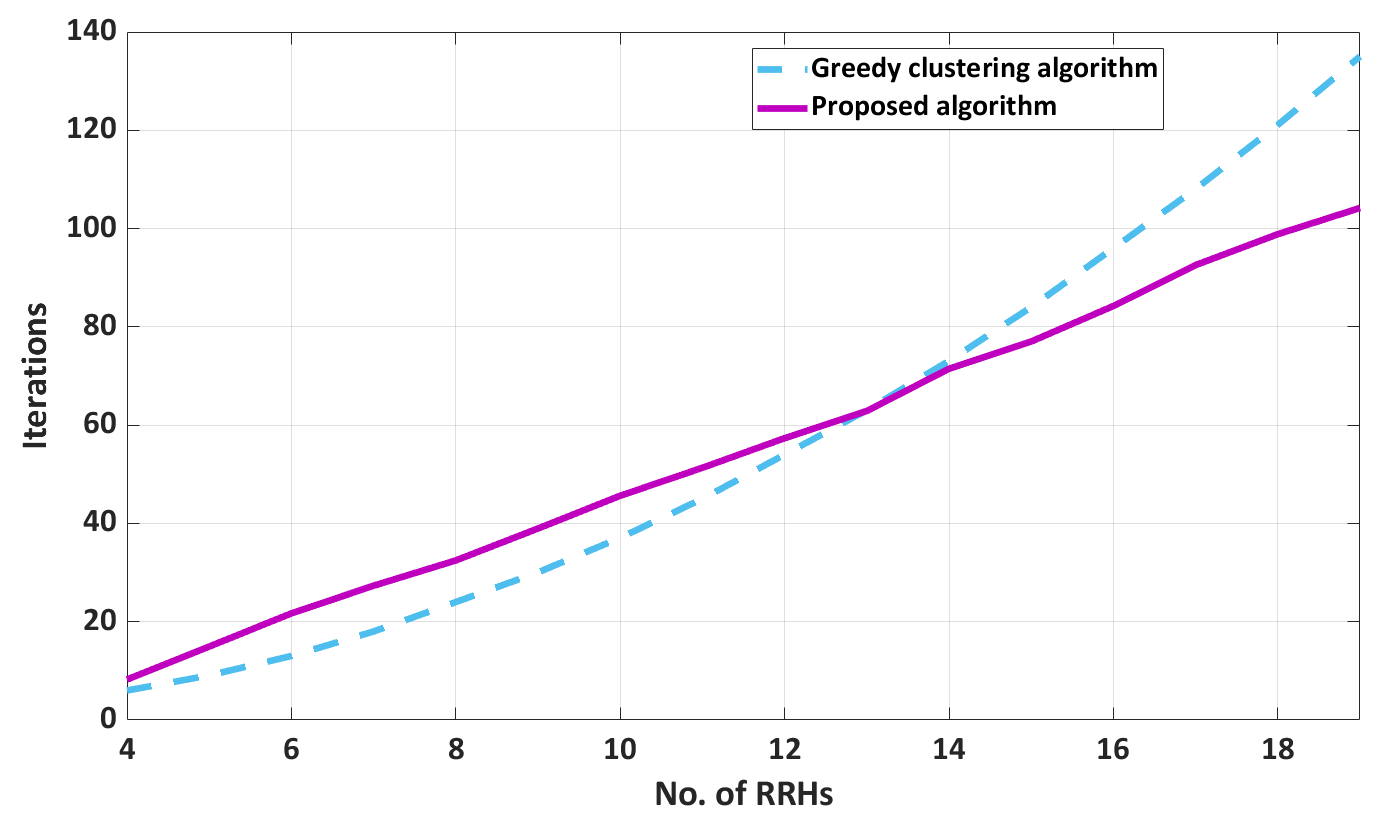}}
\caption{Average iterations versus the number of RRHs in the network.}
\label{fig14}
\end{figure}

Finally, Fig.~\ref{fig14} depicts the average iterations for the coalition formation algorithms of the GC JT-CoMP and game JT-CoMP schemes with variable number of RRHs. Fig.~\ref{fig14} confirms the linear complexity of the proposed algorithm. Furthermore, after a point, the average number of iterations of the greedy clustering algorithm used on GC JT-CoMP start to exceed that of the proposed algorithm's, confirming its unsuitability for large-scale scenarios.

\section{Conclusions}
In the 5G era, reliability and consistency in user performance is of great significance. The application of technologies such as NOMA and MU-MIMO can, on one hand, increase the service quality of the mobile users but, on the other hand, the users in the cell edges may suffer from severe intercell interference, especially when the transmission points are densely deployed. The implementation of JT-CoMP can address this problem. However, the users still compete for limited resources, meaning that intelligence needs to be behind the transmission point clustering in JT-CoMP. For this purpose, a coalition formation game was formulated to cluster the RRHs in a C-RAN based small cell scenario, where the users are considered mobile and capable of activating handovers and changing the user load of each cell.

The proposed scheme was able to form coalitions of RRHs and proved to be beneficial for the intercell interference cancellation for the cell-edge users, resulting into their instantaneous throughput increase. Furthermore, it guaranteed a moderate decrease in the instantaneous throughput of the non cell-edge users. Moreover, compared to other solutions with higher average coalition size, such as JT-CoMP with static clustering and greedy clustering, it provided more reliable and consistent results with lower feedback overhead and coordination costs, as in a dynamic environment with multiple handovers and status changes, it guaranteed an increase in the average throughput for most of the low SINR users, while it did not decrease significantly the average throughput of any user. Finally, the linear complexity of the proposed coalition formation algorithm was confirmed.


\begin{IEEEbiography}
[{\includegraphics[width=1in,height=1.25in,clip,keepaspectratio]{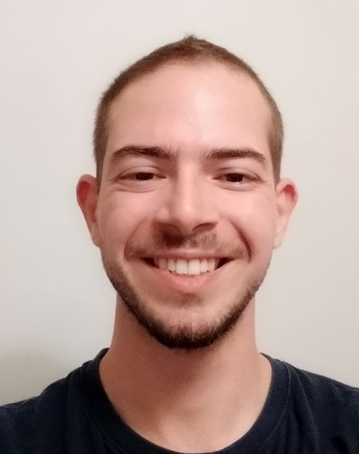}}]{\textbf{Panagiotis Georgakopoulos}} (Member, IEEE) received the Diploma degree from the Department of Electrical and Computer Engineering, University of Patras, Greece, in 2016. He is currently pursuing the Ph.D. degree with the Wireless Telecommunications Laboratory, University of Patras. He is currently a Marie Curie Early Stage Researcher with the Wireless Telecommunications Laboratory, University of Patras. His research interests include 5G and beyond networks, radio planning, radio resource management, and cooperative and self-organized networks.
\end{IEEEbiography}
\begin{IEEEbiography}
[{\includegraphics[width=1in,height=1.25in,clip,keepaspectratio]{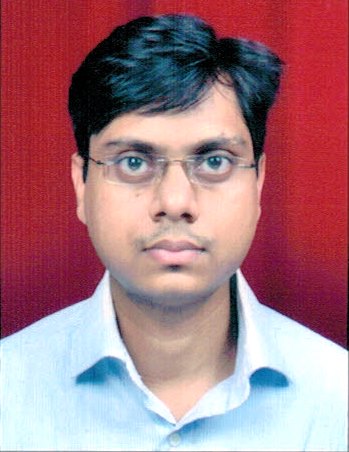}}]{\textbf{Tafseer Akhtar}} (Graduate Student Member, IEEE) received the master's degree in cyber security from the Department of Computer Engineering, National Institute of Technology of Kurukshetra, India, in 2017. He is currently an Early Stage Researcher and a Marie Curie Fellow with the Department of Electrical and Computer Engineering, University of Patras. His research interests include but are not limited to 5G small cell networks, radio resource management, game theory, distributed learning, network coded cooperation, and network security.
\end{IEEEbiography}

\begin{IEEEbiography}
[{\includegraphics[width=1in,height=1.25in,clip,keepaspectratio]{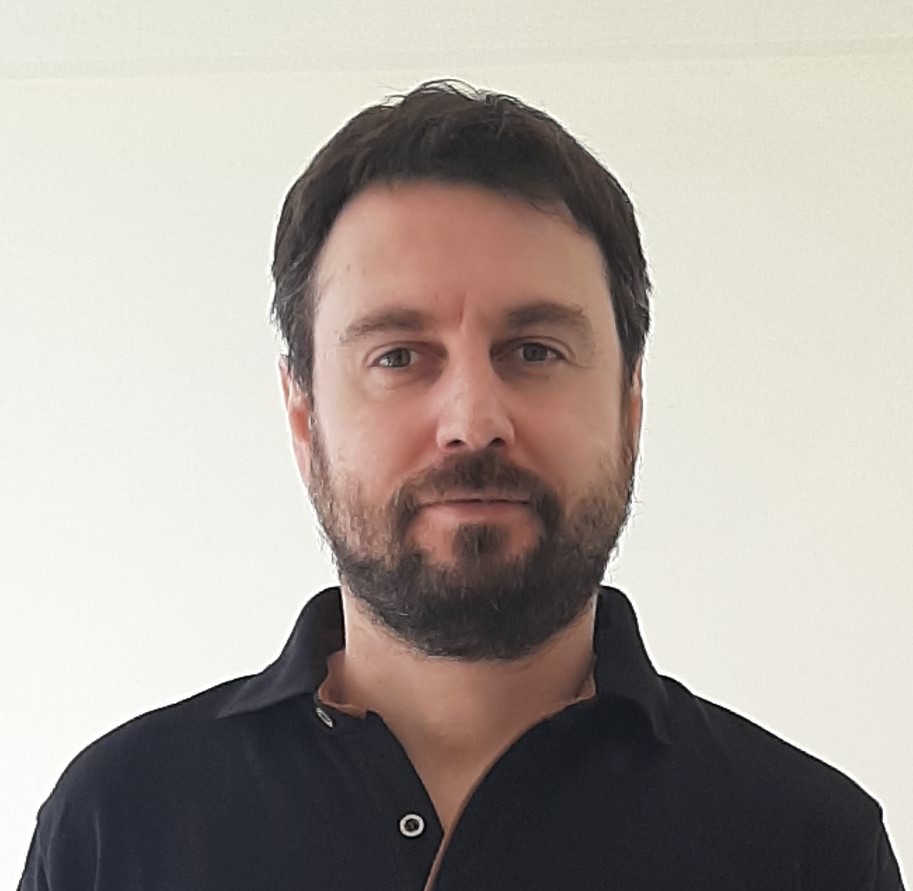}}]{\textbf{Christos Mavrokefalidis}} (Member, IEEE) received the Diploma degree in Computer Engineering and Informatics, the Masters degree in Signal Processing Systems and the PhD degree in Signal Processing for Wireless Communications, from the University of Patras, Greece, in 2004, 2006 and 2011, respectively. Since 2006, he has been a research associate with the Signal Processing and Communications Laboratory of the Computer Engineering and Informatics Department at the University of Patras, Greece. He is also affiliated (since 2006) with the Signal Processing and Communication Research Unit of the Computer Technology Institute and Press "Diophantus" at Patras, Greece and (since 2019) with the Multimedia Information Processing Systems Group of the Industrial Systems Institute of the Athena Research Center at Patras, Greece. His research interests span the scientific areas of statistical signal processing and learning with a focus on estimation theory, adaptive/distributed signal processing and sparse representations. He has been involved in numerous national, European and bilateral research projects in application areas like wireless communications and sensor networks, smart-grids and computer vision. In the past, he was also involved in the design of integrated circuits for microprocessors that support multimedia operations. Dr. Christos Mavrokefalidis is a regular reviewer in various journals and conferences in the general area of signal processing. Finally, he is a member of the Technical Chamber of Greece and IEEE.

\end{IEEEbiography}

\begin{IEEEbiography}
[{\includegraphics[width=1in,height=1.25in,clip,keepaspectratio]{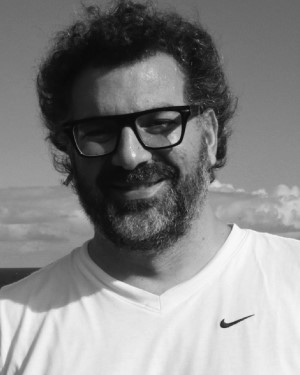}}]{\textbf{Ilias Politis}} (Member, IEEE) received the B.Sc. degree in electronic engineering from Queen Marry College, London, U.K., in 2000, the M.Sc. degree in mobile and personal communications from King’s College London, U.K., in 2001, and the Ph.D. degree in multimedia communications from the University of Patras, Greece, in 2009. He is currently a Senior Researcher at the RF, Microwave and Wireless Communications Laboratory of the University of Patras, Greece and a Senior Research Associate at the Secure Systems Laboratory of the University of Piraeus, Greece. His has served in the past as Senior Researcher with the School of Science and Technology, Hellenic Open University. He has been actively involved in all phases of several H2020 and FP7 framework projects in the areas of wireless multimedia networking, next generation networks, network security and emergency services. His research interests span over the areas of future internet and next generation networks (5G and beyond), contextual awareness, and network security, where he has published more than 90 journals and conferences.
\end{IEEEbiography}

\begin{IEEEbiography}
[{\includegraphics[width=1in,height=1.25in,clip,keepaspectratio]{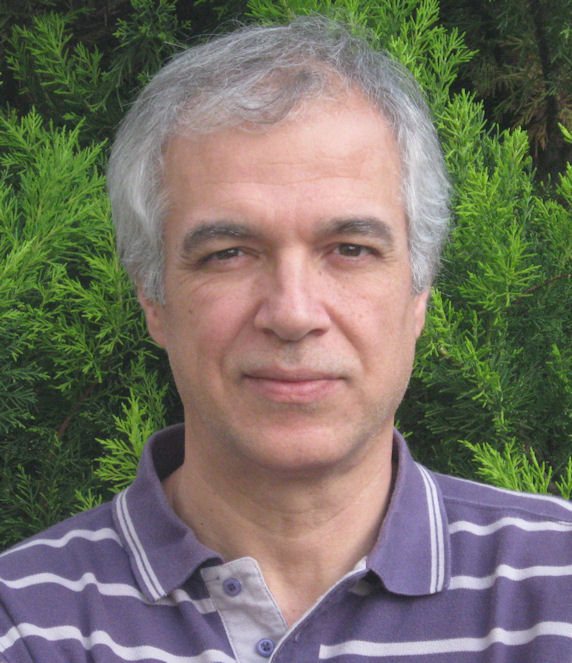}}]{\textbf{Kostas Berberidis}} (S'87-M'90-SM'07) received the Diploma degree in electrical engineering from DUTH, Greece, in 1984, and the Ph.D. degree in signal processing and communications from the University of Patras, Greece, in 1990. During 1991, he worked at the Signal Processing Laboratory of the National Defense Research Center. From 1992 to 1994 and from 1996 to 1997, he was a researcher at the Computer Technology Institute (CTI), Patras, Greece. In period 1994/95 he was a Postdoctoral Fellow at CCETT/CNET, Rennes, France. Since December 1997, he has been with the Computer Engineering and Informatics Department (CEID), University of Patras, where he is currently a Professor, and Head of the Signal Processing and Communications Laboratory.  Also, since 2008, he has been Director of the Signal Processing \& Communications Research Unit of the Computer Technology Institute and Press "Diophantus" and, since 2018, he is collaborating faculty with the Industrial Systems Institute, Athena Research Center, Greece. His research interests include distributed signal and information processing and learning, adaptive learning algorithms, signal processing for communications, wireless communications and sensor networks, array signal processing, smart grid, etc.

Prof. Berberidis has served or has been serving as member of scientific and organizing committees of several international conferences, as Associate Editor for the IEEE Transactions on Signal Processing and the IEEE Signal Processing Letters, as a Guest Editor for the EURASIP JASP and as Associate Editor for the EURASIP Journal on Advances in Signal Processing. Also, from February 2010 until December 2017 he served as Chair of the Greece Chapter of the IEEE Signal Processing Society. He has served in the “Signal Processing Theory and Methods” Technical Committee of the IEEE SPS, the “Signal Processing for Communications and Electronics” Technical Committee of the IEEE COMSOC and the EURASIP Special Area Team “Theoretical and Methodological Trends for Signal Processing”. Since January 2017 he serves as member of the Board of Directors (BoD) of EURASIP (second term started Jan. 2021). Moreover, since August 2015 he is a member of the EURASIP Technical Area Committee “Signal Processing for Multisensor Systems”. He is a member of the Technical Chamber of Greece, a member of EURASIP, and a Senior Member of the IEEE. 

\end{IEEEbiography}
\begin{IEEEbiography}
[{\includegraphics[width=1in,height=1.25in,clip]{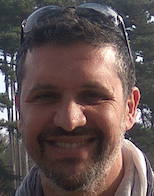}}]{\textbf{Stavros Koulouridis}} (Member, IEEE) was born in 1975, in Athens, Greece. He received the Diploma Engineer in Electrical and Computer Engineering and the Ph.D. degrees in Microwave engineering from National Technical University of Athens, Greece, in 1999 and 2003, respectively. 
From 1999 to 2003 he worked as Research Engineer in Microwave and Fiber Optics Lab and Biomedical Simulations and Imaging Unit, National Technical University of Athens. He taught several classes in the School of Pedagogic and Technological Education (ASPAITE) from 2000 to 2003. He was also teaching assistant from 2000 to 2002 in National Technical University of Athens. From 2004 to 2008, he worked as Postdoctoral Researcher at the Electroscience Laboratory, The Ohio State University, Columbus, OH, USA. In March 2009 he joined Electrical and Computer Engineering Department, University of Patras, Greece and since March 2020 he holds an Associate Professor position. Since October 2020, he is the director of RF, Microwave and Mobile Communications Laboratory. From 2015 to 2016 he was visiting Group of Electrical Engineering – Paris (GeePs) / CNRS-CentraleSupelec - Univ. Paris-Sud - Université Paris-Saclay – Sorbonne Université on a Sabbatical leave. His research interests include Antenna and Microwave Devices Design, Development and Fabrication of Novel Materials, Microwave Applications in Medicine, Electromagnetic Optimization Techniques, Applied Computational Electromagnetics. 
He was the recipient of a three year PhD Scholarship on Biomedical Engineering from Hellenic State Scholarships Foundation in 2001. In May 2005 he received the annual award for the best dissertation in National Technical University of Athens. He was the Chair of IEEE AP/MTT/ED Joint Local Greek Chapter from 2013 till 2019. He was the General Chair of IWAT 2017 (International Workshop in Antennas Technology). He has published over 100 refereed journals and conference proceeding papers. He is serving as reviewer for several scientific international journals. From 2010 till 2019 he served in the Technical Program Committee of IEEE Antennas and Propagation Society (AP-S) International Symposium. Since 2015 he serves in the Technical Program Committee of European Conference of Antennas and Propagation (EUCAP) as meta-reviewer. He is topic editor of MDPI Electronics Open Access Journal. He is Associate Editor of IEEE Antennas and Wireless Propagation Letters (AWPL) and IEEE Journal of Electromagnetics, RF and Microwaves in Medicine and Biology (J-ERM).
\end{IEEEbiography}

\EOD

\end{document}